%Paper: hep-th/9301097
%From: ZWIEBACH@IRENE.MIT.EDU
%Date: Sat, 23 Jan 1993 13:27:19 -0500 (EST)
%Date (revised): Wed, 17 Feb 1993 14:17:51 -0500 (EST)

%%%%%%%%%%%%%%%%%%%%%%%%%%%%%%%%%%%%%%%%%%%%%%%%%%%%%%%%%%%%%%%%%%%%%
% This paper refers to two figures (not included). Hard copies of
% the figures will be mailed or faxed upon request to
% grove@mitlns.bitnet (specify figures for Hata-Zwiebach).
%%%%%%%%%%%%%%%%%%%%%%%%%%%%%%%%%%%%%%%%%%%%%%%%%%%%%%%%%%%%%%%%%%%%%
\input phyzzx.tex
%%%%%%%%%%%%%%%%%%%%%%%%%%%%%%%%%%%%%%%%%%%%%%%%%%%%%%%%%%%%%%%%%%%%%
%%%%%%%%%%%%%%%%%%%%%%%%%%%%%%%%%%%%%%%%%%%%%%%%%%%%%%%%%%%%%%%%%%%%%
%
% This will make your phyzzx pagesize wider and longer
% IT IS OPTIONAL
%
\catcode`\@=11 % This allows us to modify PLAIN macros.
\def\papers{\papersize\headline=\paperheadline\footline=\paperfootline}
\def\papersize{\hsize=40pc \vsize=53pc \hoffset=0pc \voffset=1pc
   \advance\hoffset by\HOFFSET \advance\voffset by\VOFFSET
   \pagebottomfiller=0pc
   \skip\footins=\bigskipamount \normalspace }
\catcode`\@=12 % at signs are no longer letters
\papers
%
%%%%%%%%%%%%%%%%%%%%%%%%%%%%%%%%%%%%%%%%%%%%%%%%%%%%%%%%%%%%%%%%%%%%%
%%%%%%%%%%%%%%%%%%%%%%%%%%%%%%%%%%%%%%%%%%%%%%%%%%%%%%%%%%%%%%%%%%%%%

\def\square{\kern1pt\vbox{\hrule height 1.2pt
            \hbox{\vrule width 1.2pt\hskip 3pt
            \vbox{\vskip 6pt}\hskip 3pt\vrule width 0.6pt}
            \hrule height 0.6pt}\kern1pt}

\def\bra#1{\langle #1 |}
\def\ket#1{| #1 \rangle}

\def\l{{\bigl[}}
\def\r{{\bigr]}}
\def\A{{\cal A}}
\def\L{{\cal L}}
\def\O{{\cal O}}

\def\V{{\cal V}}
\def\M{{\cal M}}
\def\P{{\cal P}}

\def\e{{\epsilon}}
\def\E#1{\varepsilon(#1)}
\def\bigcdot{\hbox{\vbox{\vskip-0.1cm\hbox{\seventeenrm .}\vskip0.1cm}}}
\def\Half{{1\over 2}}
\def\wt#1{\widetilde{#1}}
\def\TV#1{{\overleftarrow{\partial}\over \partial z^{#1}}}
\def\tv#1{\overleftarrow{\partial}\!\!/\partial z^{#1}}
\def\dr#1{\overleftarrow{\partial}_{\!\! #1}}
\def\dl#1{\overrightarrow{\partial}_{\!\! #1}}
\def\p{\partial}
\def\d{{\rm d}}
\def\L{{\cal L}}
\def\O{{\cal O}}
\def\D{{\cal D}}
\def\Diff#1#2{{\partial #1\over\partial #2}}
\def\diff#1#2{\partial #1/\partial #2}
\def\DDiff#1#2#3{{\partial^2 #1\over\partial #2\partial #3}}

\def\de{\delta_\epsilon}
\baselineskip 13pt plus 1pt minus 1pt
\overfullrule=0pt
\normaldisplayskip = 20pt plus 5pt minus 10pt

%%%%%%%%%%%%%%%%%%%%%%%%%%%%%%%%%%%%%%%%%%%%%%%%%%%%%%%%%%%%%%%%%%%%%
\pubnum{MIT-CTP-2184\cr hep-th/9301097}
\date{January 1993}
\titlepage
\title{DEVELOPING THE COVARIANT BATALIN-VILKOVISKY \break
\break APPROACH TO STRING THEORY}
\author{Hiroyuki Hata \foot{On leave from
Department of Physics, Kyoto University, Kyoto 606, Japan.}
and Barton Zwiebach
\foot{Supported in part by D.O.E. contract
DE-AC02-76ER03069 and NSF grant PHY91-06210.} }
\address{Center for Theoretical Physics \break
Laboratory for Nuclear Science \break
and Department of Physics\break
Massachusetts Institute of Technology \break
Cambridge, Massachusetts 02139, U.S.A.}

\abstract{We investigate the variation of the string field action
under changes of the string field vertices giving rise
to different decompositions of the moduli spaces of Riemann surfaces.
We establish that any such change in the string action arises
from a field transformation {\it canonical} with respect to the
Batalin-Vilkovisky (BV) antibracket, and find the explicit form
of the generator of the infinitesimal transformations. Two theories
using different decompositions of moduli space are shown to yield the
same gauge fixed action upon use of different gauge fixing conditions.
We also elaborate on recent work on the covariant BV formalism, and
emphasize the necessity of a measure in the space of two dimensional
field theories in order to extend a recent analysis of background
independence to quantum string field theory.}

\endpage

\singlespace

%%%%%%%%%%%%%%%%%%%%%%%%%%%%%%%%%%%%%%%%%%%%%%%%%%%%%%%%%%%%%%%%%%%%%
\REF\batalinvilkovisky{I. A. Batalin and G. A. Vilkovisky, Phys.
Rev. {\bf D28} (1983) 2567.} %% 176
\REF\thorn{C. B. Thorn, ``String field theory'',
Phys. Rep. {\bf 174}  (1989) 1.} %% 177
\REF\hata{H.~Hata, ``BRS Invariance and Unitarity in Closed String
Field Theory", Nucl. Phys. {\bf B329} (1990) 698;
``Construction of the Quantum Action for Path-Integral Quantization
of String Field Theory", Nucl. Phys. {\bf B339} (1990) 663.} %% 177
\REF\zwiebachlong{B. Zwiebach, ``Closed String Field Theory: Quantum
Action and the BV Master Equation'', IASSNS-HEP-92/41, June 1992,
hep-th/9206084. To appear in Nucl. Phys. B.} %% 177
\REF\witten{E. Witten, ``On Background Independent Open-String
Field Theory'', IASSNS-HEP-92/53, hep-th/9208027, August 1992;\nextline
``Some Computations in Background Independent
Open-String Field Theory'', IASSNS-HEP-92/63, hep-th/9210065.}%% 204
\REF\schwarz{A. Schwarz, ``Geometry of Batalin-Vilkovisky
quantization'', UC Davis preprint, hep-th/9205088, July 1992.}%% 215
\REF\schwarzi{A. Schwarz, ``Semiclassical approximation
in Batalin-Vilkovisky formalism'', UC Davis preprint, hep-th/9210115,
October 1992.} %% 215
\REF\wittenab{E. Witten, ``A Note on the Antibracket Formalism', Mod.
Phys. Lett. {\bf A5}(1990) 487. } %% 239
\REF\brusteinalwis{R. Brustein and S. P. De Alwis, ``Renormalization
group equation and non-perturbative effects in string theory'',
Nucl. Phys {\bf B372} (1991) 451.} %% 287
\REF\sen{A. Sen, Nucl. Phys. {\bf B345} (1990) 551;
{\bf B347} (1990) 270.}
\REF\wittenzwiebach{E.~Witten and B.~Zwiebach,
``Algebraic Structures and Differential Geometry in Two-Dimensional
String Theory", Nucl. Phys. {\bf B377} (1992) 55.} %% 309

%%%%%%%%%%%%%%%%%%%%%%%%%%%%%%%%%%%%%%%%%%%%%%%%%%%%%%%%%%%%%%%%%%%%%
\chapter{Introduction and Summary}

In current formulations of string field theory the first step
in the construction consists of choosing a conformal field theory.
This conformal field theory defines a vector space, the
state space of the theory, spanned by all the (normal ordered)
local operators in the theory. A string {\it field} is simply an
arbitrary vector in this state space, and a string field theory
action is a function that given a string field gives us a
number. This action is written as a power series in the string
field and $\hbar$, and each term of the series is defined
precisely without problems of divergences or regularization.

In addition to choosing a conformal field theory, in order to
write a string field theory one must also find a way of breaking
up the moduli spaces of Riemann surfaces. This breakup is necessary
since the string amplitudes should be obtained by Feynman rules,
and these rules break the amplitudes into a sum of diagrams.
A consistent choice of string vertices and propagator gives
rise to one particular way of decomposing moduli space.
Since the physical observables are the same regardless of the
chosen vertices, it has long been thought that the possibility
of using different decompositions of moduli space must correspond
to some type of string field symmetry.

The central objective in this paper is the study of this
freedom in the choice of decomposition of moduli space
in the context of the Batalin-Vilkovisky (BV) [\batalinvilkovisky]
approach to closed string field theory [\thorn,\hata,\zwiebachlong].
As one changes the string vertices, and as a consequence, the moduli
space decomposition, the string field action changes in such a way
that the new action still satisfies the master equation.
Since the physics does not change, one suspects that the relation
between two actions using different decompositions ought to arise from
string field transformations (or redefinitions). In the BV approach
the result is actually much stronger: the relation between the two actions
arises from field transformations that are {\it canonical with respect
to the antibracket}. We give the explicit form of the generator
of these transformations. While the change of decomposition is not a gauge
invariance (the action changes), it is not so different from one;
we show that two actions using different decompositions agree
as off-shell gauge fixed actions if we use different gauge fixing
conditions.

We also discuss the use of the covariant BV formalism
in the problem of background independence of string field theory.
The conformal field theory required to write
the string field theory may be thought to be
a special point in a hypothetical space $M$ of
two-dimensional field theories.
In a recent paper Witten [\witten] has discussed how to
write a string field action $S(M)$ on a theory space $M$
(suitable for open string theory) by setting up
a BV structure on $M$ and defining an
anticommuting vector field $V_S$ satisfying $V_S^2=0$, that
plays the role of hamiltonian vector field for the function $S$.
This action, by construction, satisfies the classical BV
master equation [\batalinvilkovisky]. If such an approach is to
be eventually completed and extended to closed strings, one must
construct an action that satisfies the full quantum BV equation.
To investigate this point we make extensive use of the recent work of
A. Schwarz [\schwarz ,\schwarzi] on the geometry of BV quantization.
We argue that, in addition to the closed, non-degenerate
symplectic two-form $\omega$ in the theory space $\M$, one
also needs a suitable measure in theory space. With such measure one can
define the divergence of vector fields on supersymplectic manifolds
and the delta operator of Batalin and Vilkovisky [\schwarz].
The $V_S^2=0$ equation is generalized in such a way that the
action $S$ satisfies the full master equation.

Let us now describe briefly the contents of this paper. In \S2 we begin
by discussing the general covariance of the BV formalism,
and give a complete treatment of the differential geometry in the
appropriate supersymplectic case. We show how most of the formalism
can be developed without having to require the nilpotency of the
delta operator $\Delta_\rho$ of Batalin and Vilkovisky.
Our remarks here are mostly elaborations on the results of [\schwarz].

We then turn in \S3 to the symmetries of the master equation.
Our main objective is clarifying the physical significance of
the transformations $\delta_\e S = \Delta \e + \{S,\e\}$, which
given some action $S$ satisfying the master equation,
gives us another action
$S + \delta_\e S$ which also satisfies the master equation [\wittenab].
We emphasize that this is not a gauge symmetry and prove, within
the covariant BV approach to gauge fixing, that the observables of the
theory are not changed. As a formal development, we find an `actional'
$\A (S)$ for the `field' $S$ on the
supermanifold whose equation of motion is the master equation for $S$
and whose gauge symmetry is the above transformation $\delta_\e$.
For the classical master action we note that the above transformation
corresponds simply to a redefinition of the fields and antifields
induced by a canonical transformation. For the quantum action the
transformation $\de$ is not a field redefinition; the action also
changes due to effects having to do with the measure.
Interestingly, the original action and the perturbed one
define identical off-shell theories upon gauge fixing, if we use
different gauge fixing conditions. Using the canonical origin
of the transformations $\de$ we are able to obtain the finite
version of these transformations.

Section 4 deals with the main issue in this paper, for which
\S3 was a preparation. We show that the change in the quantum string
field master action induced by an infinitesimal change in the
decomposition of moduli space corresponds to a symmetry
transformation $\delta_\e$ for an appropriate parameter $\e$.
The parameter $\e$ has a very simple form; it is essentially obtained
by integrating a differential form over the region interpolating
between the original and final string vertices.
We find it noteworthy that changing the decomposition
of moduli space has a natural description in the BV approach to
string theory.

Finally, in \S5 we show how to generalize the formal setup
of Ref.[\witten] in order to incorporate the full quantum
master equation into the analysis. We show that the equation
$V_S^2 =0$ for the hamiltonian vector field $V_S$ arising
from the action $S$ must be changed into the equation
$V_S^2 =-\hbar V_{\hbox{div}V_S}$, and explain why it guarantees
that $S$ satisfies the quantum master equation.

\subsection{A Speculative Interpretation} The simplest
possible variation one can do to the decomposition of moduli
space of closed string field theory corresponds to changing
the length of the `stubs' in the string vertices. Since the
stub length plays the role of a cutoff in string field theory,
it has been argued that a change of stub length corresponds to
a renormalization group transformation of string field theory
[\brusteinalwis]. We are thus led to interpret the changes of
the string field action due to general variations of the moduli space
decompositions as generalized renormalization group flows.
In fact, any variation $\de S$ may be formally thought of as a
renormalization group transformation;
the action is changed, but the physics is not.
The $\de$ transformations generate a nonabelian Lie algebra, in fact,
the Lie algebra of the antibracket. The $\de$ transformations arising
from changes of moduli space decompositions do not seem to form a
subalgebra of the complete Lie algebra of $\de$ transformations.
This fact complicates a straightforward geometrical interpretation of
the nonabelian structure. Further investigation may clarify whether
or not there is a sensible notion of nonabelian renormalization
group transformations in string field theory.

\subsection{Some remarks}
We believe our results are further evidence of the surprising
efficiency of the BV formalism in dealing with string theory.
We have given technical tools that may find application in
the study of background independence, where string field redefinitions
are often necessary [\sen]. Our analysis may also help find
gauge fixing conditions different from those of the Siegel gauge; such
gauges seem necessary for cases when the closed string semirelative
cohomology is nontrivial, as is the case in $c=1$ strings
[\wittenzwiebach]. We have clarified the sense in which
the choice of moduli space decomposition corresponds to a symmetry
and has a special role in the BV formalism.
If one wished to promote these transformations to {\it gauge}
invariances, the only option we see is that of extending the number
of dynamical variables in the string field theory.
Finally, our discussion has emphasized the need for a suitable density
$\rho$ in the space of 2d field theories leading to a nilpotent
$\Delta_\rho$.

\REF\lianzuckerman{B. H. Lian and G. Zuckerman,
``New Perspectives on the BRST-algebraic structure of string theory'',
hep-th/9211072, Yale preprint, November 1992.} %% 323
\REF\batalintyutin{I.A.~Batalin and I.V.~Tyutin, ``On Possible
Generalizations of Field-Antifield Formalism",
Lebedev Inst. preprint, FIAN/TD/18-92, hep-th/9211096.} %% 323
\REF\getzler{E.~Getzler, ``Batalin-Vilkovisky Algebras and
Two-dimensional Topological Field Theories",
Dept. Math. MIT preprint, hep-th/9212043.} %% 324
\REF\schwarzpenkava{M.~Penkava and A.~Schwarz,
``On Some Algebraic Structures Arising in String Theory",
UC Davis preprint, UCD-92-03, hep-th/9212072.} %% 324

\noindent
There have been several recent works on Batalin-Vilkovisky theory.
Some of our results in \S2 have also been obtained by Lian and Zuckerman
[\lianzuckerman], Batalin and Tyutin [\batalintyutin],
Getzler [\getzler], and, Schwarz and Penkava [\schwarzpenkava].

\REF\henneaux{M. Henneaux, `Lectures on the Antifield-BRST formalism
for gauge theories', Proceedings of the the XXI GIFT International Seminar
on Theoretical Physics.}
\REF\henneauxteitelboim{M. Henneaux and C. Teitelboim, `Quantization
of Gauge Systems', Princeton University Press, Princeton, New Jersey,
1992.} %% 333
\REF\alfaro{J. Alfaro and P.H. Damgaard, `Field transformations,
collective coordinates and BRST invariance', Ann. Phys. {\bf 202} (1990) 398;
`BRST symmetry of field redefinitions', Ann. Phys. {\bf 220} (1992) 188.}
\REF\voronovtyutin{B. L. Voronov and I.V. Tyutin, `Formulation of
gauge theories of general form'. Theor. Math. Phys.
{\bf 50} (1982) 218.}

\noindent
M. Henneaux has brought to our attention that the $\de$ transformations
were familiar in the context of the ambiguities of the solution
of the master equation [\henneaux,\henneauxteitelboim]. The role
of field transformations (canonical or more general) in the BRST
formalism has also been studied in works of Alfaro and Damgaard
[\alfaro ]. Finally the connection between canonical transformations
and change of gauge conditions was understood in the context of the
`Zinn-Justin' equation by Voronov and Tyutin [\voronovtyutin ].

\REF\hikkop{H.~Hata, K.~Itoh, T.~Kugo, H.~Kunitomo, and K.~Ogawa, %%547
Phys. Rev. {\bf D35} (1987) 1318; Nucl. Phys. {\bf B283} (1987) 433.}
\REF\sonodazwiebach{H. Sonoda and B. Zwiebach, Nucl. Phys.
{\bf B331} (1990) 592.} %% 1106

%%%%%%%%%%%%%%%%%%%%%%%%%%%%%%%%%%%%%%%%%%%%%%%%%%%%%%%%%%%%%%%%%%%%%
\chapter{The BV equation on a general coordinate system}

In this section we will discuss the covariant formulation of the
Batalin-Vilkovisky formalism, and summarize basic results needed for
this paper.\foot{For an introduction to BV theory and a review of
earlier developments see [\henneauxteitelboim].}
We were motivated by the recent discussion of A. Schwarz [\schwarz],
that provided, in our opinion, the main insight into the covariant
formulation. His proposal is that the supersymplectic manifold
of fields and antifields, carrying the symplectic form $\omega$, must
be endowed with a volume element $\mu$, or equivalently, a density
function $\rho$.
This allows one to define the so-called delta operator $\Delta_\rho$
of Batalin and Vilkovisky as the second order differential
operator that acting on a function gives the divergence of the vector
field arising from that function. The volume element is necessary to
define the divergence. One then imposes the
condition $\Delta_\rho^2 =0$ and shows that it leads to a sensible
formalism.

In this section we will
show that the standard formulas in the formalism
{\it hold for arbitrary} $\rho$. That is, without imposing
the $\Delta_\rho^2=0$ condition, we verify that the antibracket
(which is $\rho$-independent) measures the failure of $\Delta_\rho$
to be a derivation of pointwise multiplication, and $\Delta_\rho$
satisfies the Leibniz rule on the antibracket. We also note that
$\Delta_\rho^2$,
which naively would be expected to be a fourth order differential
operator, is actually a first order operator. We conclude that the
nilpotency condition of $\Delta_\rho$ is (at present) imposed only
because it is clear that, with Darboux coordinates {\it and} $\rho=1$,
consistent quantization can be done, and the recent work of
Schwarz shows that the nilpotency  of $\Delta_\rho$ insures that
such a preferred Darboux system of
coordinates can be found.\foot{General (super)canonical transformations,
defined to preserve the Darboux form, do not preserve the measure
factor $\rho$.}

Consider a $(n, n)$-dimensional supermanifold $\M$ of fields
and antifields.
This supermanifold is endowed with an odd symplectic structure defined
by an odd two-form $\omega$ which is non-degenerate and
closed, $\d\omega=0$.
In a local coordinate system $(z^I)=(z^1,z^2,\ldots,z^{2n})$,
$\omega$ is given by\foot{In Appendix A we summarize the rules of
exterior calculus on a supermanifold, and in Appendix B we give
some properties of the basic ingredients of the formalism.}
$$
\omega = - \d z^I \omega_{IJ}(z) \d z^J
= \omega_{JI}(z)\d z^I\wedge \d z^J\ .
\eqn\OMEGA
$$
In familiar applications of the BV formalism one adopts Darboux
coordinates where $\omega$ takes the form
$\omega= -2\, \d \phi^i\wedge\d \phi^*_i$ ($\phi^i$ and $\phi^*_i$
are fields and antifields, respectively).
However, since we are interested in the {\it covariant}
aspects of the BV formalism, we will consider general coordinate
systems.

In correspondence to the Poisson bracket in bosonic symplectic
manifolds, we have the antibracket defined by the
inverse matrix $\omega^{IJ}$ as
$$
\{A,B\} \equiv A\dr{I}\, \omega^{IJ} \,\dl{J}B
= (-)^{I(A+1)}\p_I A\bigcdot \,\omega^{IJ}\, \bigcdot\p_J B\ ,
\eqn\ANTIBRACKET
$$
where $\dl{I}=\p_I=\p_l/\p z^I$ and $\dr{I}=\p_r/\p z^I$ denote
the left- and right-derivatives, respectively.
It follows from the above that the antibracket is also given by
$$
\{A,B\} = \omega(V_A,V_B) = V_B(A)\ ,
\eqn\ANTIBRACKETII
$$
where $V_A$ is the hamiltonian vector field corresponding to a
function $A$:
$$
V_A = \TV{I}\,\omega^{IJ}\p_J A\ .
\eqn\HV
$$
Just as in the case of ordinary symplectic manifolds, $\d\omega=0$
implies that the antibracket satisfies a (graded) Jacobi identity:
$$
(-)^{(A+1)(C+1)}\bigl\{\{A,B\},C\bigr\} + \hbox{cyclic}(A,B,C) = 0\ ,
\eqn\JACOBI
$$
and the same equation with $\bigl\{\{A,B\},C\bigr\}$ replaced
by $\bigl\{A,\{B,C\}\bigr\}$.

Defining the Lie-bracket $[V,W]$ of two vector fields
$V=(\tv{I})V^I$ and $W=(\tv{I})W^I$ by
$$
[V,W] \equiv VW - (-)^{VW}WV
= \TV{I} \left(V^I\dr{J}W^J -(-)^{VW}W^I\dr{J}V^J\right)\ ,
\eqn\LIEBRACKET
$$
the condition $\d\omega=0$ also implies that hamiltonian vector
fields form a Lie subalgebra of the Lie algebra of vector fields
$$
[V_A,V_B]=V_{\{A,B\}}\ .
\eqn\ALGEBRA
$$
This is the standard formula establishing the homomorphism
from the Lie algebra of the antibracket to the Lie algebra of
vector fields (it is not an isomorphism because all the constant
functions are mapped to the zero vector).

We now note that since the supersymplectic form $\omega$ is odd
we have $\omega \wedge \omega = 0$, and, in contrast to ordinary
symplectic theory, we cannot raise $\omega$ to some power in order
to obtain a canonical volume form. In other words there is no a-priori
``Liouville'' volume element.\foot{Recall that in supermanifolds
the volume element is not a differential form.} A related fact
is that in ordinary symplectic theory a hamiltonian vector field $V_f$
(associated to a function $f$)
preserves the symplectic form: $\L_{{}_{V_f}} \omega =0$.
As a consequence
$\L_{{}_{V_f}} \mu =0$ ($\mu \sim \omega^n$),
and via the standard relation
$\mu \bigcdot (\hbox{div}_\mu V ) = \L_{{}_V} \mu$,
we obtain the result that hamiltonian vector fields are divergenceless.
In the supersymplectic case hamiltonian
vector fields need not be divergenceless.

Following Ref.[\schwarz] we introduce a volume element by
$$
d\mu(z) = \rho(z)\prod_{I=1}^{2n}\d z^I \ .
\eqn\MU
$$
where $\rho(z)$ is a density.
Then, defining the divergence of a vector field $V=(\tv{I})V^I$
by
$$
\hbox{div}_\rho V = {1\over \rho}(-)^I \p_I\left(\rho V^I\right)\ ,
\eqn\DIV
$$
we can finally introduce the second order differential operator
$\Delta_\rho$ of the BV formalism as an operator that acts on
functions to give functions:
$$
\Delta_\rho A \,\equiv \, \Half \, \hbox{div}_\rho V_A
= {1\over 2\rho}(-)^I \p_I\left(\rho\omega^{IJ}\p_J A\right)\ .
\eqn\DELTA
$$
Namely, $\Delta_\rho$ acting on a function gives
the divergence of the hamiltonian vector field associated to the
function.
Note that $\Delta_\rho$ is odd, \ie, $\E{\Delta_\rho}=1$.
It is easily seen that two $\Delta$'s corresponding to two different
$\rho$'s are related by
$$
\Delta_{\wt{\rho}} A = \Delta_\rho A
+ \Half \, \{ \, \ln (\tilde\rho / \rho ), A \} \ .
\eqn\DELTARHOTILDE
$$
There is a very simple relation between the delta operator and
the antibracket [\wittenab]; the antibracket measures the failure
of delta to be a derivation of the associative algebra of functions
(under pointwise multiplication). This actually holds for
arbitrary $\rho$:
$$
(-)^A \{A,B\} = \Delta_\rho(A\bigcdot B) - \Delta_\rho A\bigcdot B
- (-)^A A\bigcdot \Delta_\rho B \ ,
\eqn\ABDELTA
$$
since the $\rho$-dependence on the right hand side cancels out.
An important property of $\Delta_\rho$ is the Leibniz rule for the
antibracket,
$$
\Delta_\rho\{A,B\} = \{\Delta_\rho A, B\}
+ (-)^{A+1}\{A, \Delta_\rho B\}\ .
\eqn\LEIBNIZ
$$
Again, we emphasize that Eqn.\LEIBNIZ\ holds for any $\rho$.
This is proven using Darboux coordinates as follows.
One adopts Darboux coordinates and denotes
the new density by $\tilde \rho$.
One then verifies that in those Darboux coordinates $\Delta_{\rho=1}$
does satisfy the Leibniz rule \LEIBNIZ. The Leibniz rule for
$\Delta_{\wt{\rho}}$ then follows from  Eqn.\DELTARHOTILDE\ and
the Jacobi identity \JACOBI.
In fact, using \ABDELTA\ repeatedly we can show that
$$\eqalign{\Delta_\rho\{A,B\} = &\{\Delta_\rho A, B\}
+ (-)^{A+1}\{A, \Delta_\rho B\} \cr
&+ (-)^A\left[ \Delta_\rho^2(A\bigcdot B)
- (\Delta_\rho^2 A)\bigcdot B
- A\bigcdot(\Delta_\rho^2 B)\right]\ ,\cr}\eqn\CALLEIBNIZ$$
and therefore the Leibniz rule for the antibracket established
above requires that $\Delta_\rho^2$ should satisfy
$$\Delta_\rho^2(A\bigcdot B) = (\Delta_\rho^2 A)\bigcdot B
+ A\bigcdot(\Delta_\rho^2 B) \ .\eqn\DELTASQLEIBNIZ$$
This implies that $\Delta_\rho^2$, which is naively
a fourth order differential operator, is in fact a
{\it first order} differential operator for any $\rho$,
$$\left(\Delta_\rho\right)^2 = \Half\left[\Delta_\rho
{1\over\rho}(-)^I\p_I\left(\rho\omega^{IJ}\right)\right]
\!\bigcdot\,\p_J\ .\eqn\DELTASQUARE$$
Equation \DELTASQUARE\ (or \DELTASQLEIBNIZ) can also be proved directly
by a tedious but straightforward calculation using $\d\omega=0$.
Turning this around we see that the condition $\Delta_\rho^2=0$
would imply, from the vanishing of the third order
differential part, the condition $\d\omega=0$. This has
been found independently in Refs.[\schwarzpenkava ,\getzler].
The Jacobi identity \JACOBI, the Leibniz rule \LEIBNIZ\ and
the nilpotency condition $\Delta_\rho^2=0$ (to be imposed below)
are the fundamental properties of the antibracket $\{\cdot,\cdot\}$
and $\Delta_\rho$. Note the formal resemblance of $\Delta$ and the
antibracket $\{\cdot,\cdot\}$
to the BRST operator $Q$ and the $*$-product of string fields
in the cubic light-cone-style HIKKO classical closed string field
theory [\hikkop]. The latter $(Q, *)$, also
satisfy the conditions of nilpotency, Leibniz rule and Jacobi identity
(the analog of pointwise multiplication has not been found necessary
to write the string field theory.).

For the particular case of Darboux coordinates and $\rho=1$,
the antibracket and the delta operator,
with $(-)^i\equiv(-)^{\E{\phi^i}}$, read
$$\eqalign{&\{A,B\} = \Diff{_r A}{\phi^i}\Diff{_l B}{\phi^*_i}
- \Diff{_r A}{\phi^*_i}\Diff{_l B}{\phi^i} \ , \cr
&\Delta = (-)^i \Diff{_l}{\phi^i}\Diff{_l}{\phi^*_i} \ , \cr}
\eqn\ABDINDB$$

Having given all the necessary notation,
the quantum BV equation for the master action $S(z)$ is given by
$$\Delta_\rho\, e^{S(z)/\hbar} = 0\ ,\eqn\BVI$$
or equivalently
$$\hbar\Delta_\rho S + \Half\{S,S\} = 0\ .\eqn\BVII$$
Under general coordinate transformations in the supersymplectic
manifold $\wt{z}^I=\wt{z}^I(z)$ (preserving Grassmanality:
$\E{\wt{z}^I}=\E{z^I}=I$), the action $S$, the antibracket,
and $\Delta$ transform as scalars,
while $\wt{\omega}^{IJ}$ and $\wt{\rho}$ in the new coordinate
system are given by
$$\eqalignno{&\wt{\omega}^{IJ}(\wt{z}) = \{\wt{z}^I,\wt{z}^J\}
= \wt{z}^I\dr{K}\,\cdot\omega^{KL}(z)\cdot\,\dl{L}\wt{z}^J\ ,
&\eqname{\OMEGATILDE} \cr
\noalign{\vskip0.5cm}
&\wt{\rho}(\wt{z}) = \rho(z)\bigcdot\exp\left\{ - \hbox{sTr}\ln\left(
{\p_l \wt{z}^J \over \p z^I }\right)\right\}
= \rho(z)\bigcdot\hbox{sdet}\!
\left( {\p_l z^J \over \p\wt{z}^I } \right)\ ,&\eqname{\RHOTILDE} \cr}$$
where $\hbox{sTr}(M_I^J)\equiv (-)^I M_I^I$.

In BV quantization the density function $\rho$ defining the
operator $\Delta_\rho$ is not arbitrary. One imposes the
condition of {\it nilpotency}:
$$\left(\Delta_\rho\right)^2 = 0\ .\eqn\NILPOTENCY$$
(Recall that $\Delta_\rho^2$ is in general a first order differential
operator.)
This condition is necessary and sufficient to prove that there exists
a Darboux frame with $\rho =1$ [\schwarz]. The existence of such a frame
is sufficient to guarantee, by the standard argument of
Ref.[\batalinvilkovisky], consistent path-integral
quantization using the master action $S$.

Given a solution $S(z)$ of the BV equation \BVII,
the quantization is given as follows [\schwarz].
First, one chooses a lagrangian submanifold $L$, {\it i.e.}, a
$(k,n-k)$-dimensional submanifold of $\M$, such that
$\omega(v,\tilde v) = 0$ for any pair, $v$ and $\tilde v$, of
vectors tangent to $L$ at $z$ ($v, \tilde v \in T_z L$).
The choice of $L$ corresponds to the choice of gauge fixing.
The observables are defined by integrals over the lagrangian
submanifold. This requires a volume element $d\lambda$ on $L$,
that can be obtained canonically using the volume element $d\mu$ on
the whole supermanifold, and the two form $\omega$. One defines
$$d\lambda (e_1,\cdots,e_n)=d\mu (e_1,\cdots,e_n,f^1,\cdots,f^n)^{1/2}
\ ,\eqn\LAMBDA$$
where $d\mu$ is the volume element in $\M$, Eqn.\MU,
and $(e_1,\ldots,e_n,f^1,\ldots,f^n)$ is a basis of the tangent
space $T_z\M$ such that $(e_1,\ldots,e_n)$ is a basis of
$T_z L$ and the condition $\omega(e_i,f^j)=\delta_i^j$ is satisfied.
Using the lagrangian submanifold $L$ and the associated volume element
$d\lambda$, the quantum theory is defined by the path-integral
$$\int_L\! d\lambda \, e^{S/\hbar} \ .\eqn\PATHINTEGRAL$$
It can be shown that \PATHINTEGRAL\ is invariant under deformations
of $L$. A convenient way to do gauge fixing
requires finding a set of $n$ linearly independent constraints $G_i =0$
in involution
$$ \{ G_i , G_j \} = U_{ij}^k \, G_k ,\eqn\invol$$
with $U$'s a set of possibly field/antifield dependent structure
constants. Condition \invol\ implies that the submanifold defined
by $G_i =0, \forall i$, will be a lagrangian submanifold. This is
seen as follows. Let $V_i$ be the hamiltonian vector associated
to $G_i$. The vector $V_i$ is tangent to the submanifold since
$V_i(G_j) = \{G_j,G_i\} =0$ on the submanifold. Thus the vectors
$V_i$ form a basis for the tangent space to the submanifold.
Then, for any two tangent vectors $V$ and $V'$, we have
$\omega (V, V') = \omega (a^i V_i , b^j V_j )$
$\sim a^ib^j\omega (V_i ,V_j) = a^ib^j \{ G_i , G_j \} = 0$, on
the submanifold.

In the conventional situation the above constraints are used to
determine the antifields as functions of the fields. This
requires that
$$G_i = \Lambda^{~j}_i(\phi ,\phi^*)\, (\phi^*_j-f_j(\phi )),
\eqn\ggferm$$
with $\Lambda$ an invertible matrix [\batalintyutin].
Eqn.\invol\ requires
that on the constraint surface $\{ G_i , G_j \} = 0$, and this implies
$$ \partial_i f_j - \partial_j f_i = 0 \quad\Rightarrow\quad
f_i = {\partial \Upsilon \over \partial \phi^i},\eqn\eeffgd$$
where $\Upsilon$ is the so-called gauge fermion ($\E{\Upsilon}=+1$).
The gauge fixing conditions therefore read
$$ \phi^*_i = {\partial \Upsilon \over \partial \phi^i}.\eqn\eeffxd$$
This concludes our presentation of the covariant form
of the Batalin-Vilkovisky formalism.

%%%%%%%%%%%%%%%%%%%%%%%%%%%%%%%%%%%%%%%%%%%%%%%%%%%%%%%%%%%%%%%%%%%%%
\chapter{Symmetries of the Master Equation}

In this section we will discuss a symmetry of the master
equation. Its origin is the well-known invariance of the classical
master equation $\{ S,S \}=0$ under the transformations
$\delta S = \{ S , \e \}$. This symmetry can be
extended to the quantum master equation as
noticed in [\wittenab] (it corresponds to an ambiguity in solving
the quantum BV master equation [\henneauxteitelboim ]).
In this section, we will show, using the covariant description
of gauge fixing, that the symmetry transformation, while changing
the master action, preserves all the observables of the theory.
We will also give an action for the master equation, invariant
under the symmetry transformation. By using the canonical origin
of the transformation we establish that two actions related by
this transformation lead to the same gauge fixed theory upon
use of different gauge fixing conditions.
Finally we present the finite version of the transformations.

The above comments can be made somewhat more precisely.
Given a supermanifold $\M$ with structure $(\omega , \rho )$,
the action $S$ is a function on $\M$. If $S$ satisfies the
master equation, the transformation gives us a new $S$ that
also satisfies the equation, without changing the structure
$(\omega , \rho )$ on $\M$. An off-shell gauge fixed theory
is defined by the pair $(S,L)$, where $L$ is the gauge fixing
surface. If the transformation takes $S\to S'$, we claim there
is an $L'$ such that $(S',L')$ is an equivalent off-shell theory.

%%%%%%%%%%%%%%%%%%%%%%%%%%%%%%%%%%%%%%%%%%%%%%%%%%%
\section{Symmetry Transformations and Observables}

The master equation $\Delta e^S =0$ (in this section we omit the
$\hbar$ dependence and the subscript $\rho$ in $\Delta$) can be
written as
$$M(S) = e^{-S} \Delta e^S = \Delta S+{1\over 2}\{ S ,S\}=0,
\eqn\meagain$$
and a solution $S$ of this equation can be used to define a
quantum field theory. In particular,
it has been observed [\wittenab] that, given a solution $S$,
one can generate other
solutions using the following infinitesimal transformation
$$\delta_\e S = \Delta \e + \{ S, \e \} ,\eqn\wobs$$
with $\e$ an odd parameter. The second term on the right hand side
is well-known; it generates a canonical transformation of the
master action and preserves the classical master equation. The
first term is a quantum correction.
Indeed, making use of $\Delta^2 = 0$ along with
the Leibniz rule of $\Delta$ acting on the antibracket,
one verifies that
$$\delta_\e M(S) = \{ M(S) , \e \}, \eqn\onever$$
which implies that if $S$ satisfies the master equation,
$S+ \delta_\e S$ will too.
Computing the commutator of two such transformation we find that
$$\l \delta_{\e_1} , \delta_{\e_2} \r =
\delta_{\{ \e_1 , \e_2 \} } .\eqn\lalgeb$$
Thus the algebra of these transformations is simply the
Lie algebra of the antibracket.

It is obvious that this transformation does not correspond to a gauge
transformation of the field theory in question; gauge transformations
should leave the action invariant, and the above manifestly does not.
The transformations change the action but one expects that the new
action defines a physically equivalent theory.
Given the results of Ref.[\schwarz] it is possible to
prove this explicitly by showing that the observables are not changed.
The simplest observable is the partition function. Under this
transformation we have that
$$\eqalign{
Z = \int_L d\lambda \, e^S \, \rightarrow
\,& \int_L\, d\lambda e^{S+\Delta \e + \{ S,\e \} } \cr
 \, &=  \int_L d\lambda e^S +
\int_L d\lambda e^S (\Delta \e + \{ S , \e \} ) + \O (\e^2) .\cr }
\eqn\modpart$$
The change in the partition function is therefore given by
$$\delta_\e Z = \int_L d\lambda \, e^S (\Delta \e + \{ S , \e \} ) =
\int_L d\lambda\, \Delta \bigl( \e e^S \bigr) = 0,\eqn\ncinz$$
where use was made of
Eqn.\ABDELTA\ together with
$\Delta  e^S = 0$,  and the final equality follows from
$\int_L d\lambda \,\Delta F =0$, which holds for any $F$, as proven in
[\schwarz] for the case of compact lagrangian submanifolds $L$.
\foot{This was proven for finite dimensional supermanifolds.
We assume it also holds for the infinite dimensional case.} This
shows that the partition function has not changed. Let us now
consider the case of other observables. Given an operator
$A$, one has an observable if the expectation value (in the
theory defined by $S$)
$$ \langle A \rangle_{{}_S} \equiv \int_L d \lambda A \, e^S,
\eqn\gtobs$$
is independent of gauge fixing (deformations of the lagrangian
submanifold $L$). This holds if
$$ \Delta \bigl( A e^S \bigr) = 0. \eqn\cobser$$
or equivalently (using $\Delta e^S=0$),
$$ \delta_A S=\Delta A + \{S, A\} = 0 . \eqn\cobserb$$
If we change the action we must also change the operator $A$ so
that it remains an observable.  This requires that
$$\Delta \bigl( (A+ \delta_\e A) e^{S+ \delta_\e S} \bigr) = 0,
\eqn\cobserc$$
and this condition implies that the variation of $A$ is given by
$$ \delta_\e A = \{ A , \e \} .\eqn\nobser$$
Then a simple computation gives
$$ \langle  A+ \delta_\e A \rangle_{{}_{S + \delta_\e S} }=
\langle A \rangle_S + \int_L d\lambda\, \Delta \bigl( \e A e^S \bigr)
= \langle A \rangle_{{}_S} , \eqn\getinv$$
as desired. This proves that the observables are the same in the
two theories, and therefore that the two theories are physically
equivalent.

%%%%%%%%%%%%%%%%%%%%%%%%%%%%%%%%%%%%%%%%%%%%%%%%%%%
\section{An Action for the Master Equation}

We have seen that the transformations \wobs\ are not
gauge transformations. Nevertheless,
since they leave the master equation invariant, one can
expect the transformations to correspond to gauge invariances
of an actional $\A(S)$ whose equation of motion is the master
equation $M(S)=0$. Such an actional exists and it is given by
$$
\A (S) = - \Half\int_{\M} d\mu \,\{e^S , e^S \}
= \int_{\M} d\mu\,e^S\Delta e^S ,
\eqn\ameee
$$
where $d\mu$ is the volume element on the supermanifold $\M$ given
by Eqn.\MU, and the second expression follows from
$$\int_{\M} d\mu\, A\, \Delta B
= - \Half\int_{\M} d\mu\,\{A,B\}
= (-)^A \int_{\M} d\mu\, (\Delta A ) B ,
\eqn\saddel$$
which in turn, results from integration by parts (dropping total
derivatives). It is clear that the equation of motion for $S$ resulting
from the actional $\A(S)$ is indeed the master equation $\Delta e^S=0$.
Gauge invariance of $\A(S)$ is shown as follows:
$$
\eqalign{
\de\A(S)&=-\int_{\M} d\mu\,\bigl\{e^S,e^S\Delta\e
+ \{e^S,\e\}\bigr\} \cr
&=-\int_{\M} d\mu\,\left(
\Half\{e^{2S},\Delta\e\} + \{e^S,e^S\}\Delta\e
+ \Half\bigl\{\{e^S,e^S\},\e\bigr\} \right) \cr
&= \int_{\M} d\mu\,e^{2S}\Delta^2\e = 0 ,\cr }\eqn\deas
$$
where the second equality is a consequence of Eqn.(B.3) and
$\bigl\{e^S,\{e^S,\e\}\bigr\}=\half\bigl\{\{e^S,e^S\}, \e\bigr\}$
(which follows from the Jacobi identity \JACOBI),
and the third equality requires the use of Eqn.\saddel.

When the supermanifold $\M$ is a fixed space of fields and antifields,
the dynamical variable $S$ represents a possible action, and
$\A (S)$ is a function on the space of actions,
having as critical points the actions that define quantum field
theories (with the given field content).

While this actional $\A(S)$ has the expected properties,
its possible significance is not clear to us.
The most naive (and radical!) interpretation of $\A(S)$ would be
as an action for a quantum theory with the dynamical variable $S$
(for an $(n,n)$ supermanifold $\M$, $\A(S)$ is even if $n$ is odd).
This ``theory of theories" would be described by a path-integral:
$$\int\!\D S\,\exp\!\left({1\over \lambda}\A(S)\right) ,\eqn\DS$$
where $\lambda$ is the coupling constant.
\foot{Although the classical theory of $\A(S)$ is a free field theory
if we take $e^S$, instead of $S$, as the fundamental variable,
the range of path-integration over the new variable $e^S$
in the quantum theory \DS\ is then non-trivial.
We thank E. Witten for his comments on this point.}
The path-integral \DS\ requires gauge fixing
of the gauge symmetry $\de$, and we would have to consider the BV
equation for the quantum action having $\A(S)$ as its classical part.
It may be of help to notice that,
similarly to the case of string field theories [\hikkop,\zwiebachlong],
$\A(S)$ also has an invariance under the pre-BRST transformation
$\delta_{\rm B}S=M(S)=\Delta S + \half\{S,S\}$, which is nilpotent,
$(\delta_{\rm B})^2=0$.

%%%%%%%%%%%%%%%%%%%%%%%%%%%%%%%%%%%%%%%%%%%%%%%%%%%
\section{Interpretation of $\de$ as a Change of Gauge Fixing}

The transformation $\de S$ of the master action (Eqn.\wobs)
and $\de A$ of the observables (Eqn.\nobser) are essentially
infinitesimal canonical transformations generated by $\e$.
The purpose of the present subsection is to find the field
theoretic interpretation of such transformations.
Since the master action does not have a direct physical interpretation,
our discussion will be done in the context of the gauge fixed theory.
Since these transformations do not change the observables of the
theory, we anticipate that two theories differing by such
a transformation
could be related by a field redefinition, and more interestingly,
by a change of gauge. We will indeed show that this is the case.

Consider the supermanifold $\M$ of fields and antifields, and
a (Grassmann-odd) function $\alpha$ with $V_\alpha$ the associated
hamiltonian vector field. Let $g_t$ be the
diffeomorphism generated by following the integral curves of
the vector field $V_\alpha$ for a parameter distance $t$.
We then have that for small $t$ the diffeomorphism takes
$z^I$ to $z^I + t V_\alpha^I +\O (t^2)$, or in other words
$$ g_t : z^I \rightarrow z^I+t\{ z^I,\alpha\}+\O (t^2),\eqn\inddif$$
where use was made of Eqn.\ANTIBRACKETII.
For any scalar $F$ we then have that
$$g_t^* F(z^I ) = F \bigl( g_t (z^I )\bigr) =
F \bigl( z^I+ t\{ z^I , \alpha \}+\O(t^2) \bigr)
= F + t \{ F , \alpha \} + \O (t^2) .\eqn\pullscal$$
For the case of the action $S$ we would have that
$$g_t^* S = S + \{ S , t\alpha \} = S + \delta_{t\alpha}^{class} S,
\eqn\varclas$$
which indicates that
the classical part of the transformation $\delta_{t\alpha} S$
(the part neglecting $\Delta$) corresponds simply
to a field redefinition implementing \inddif. Since the classical
part of the transformation defines the transformation of the
classical master action (the $\hbar =0$ limit of the master action),
we conclude that {\it the transformation of the classical master action
corresponds to a canonical field redefinition.}
The complete transformation $\delta_{t\alpha}$ of the master action
{\it does not} correspond to a canonical redefinition (even though it
arises as a consequence of one) because it includes extra
terms coming from the measure. To clarify this we must consider
gauge fixing, or, in other words, choosing a lagrangian submanifold $L$.

The diffeomorphism pushes the lagrangian submanifold $L$
to a new submanifold $L_t$ which is also lagrangian
$$g_t : L \rightarrow L_t . \eqn\movelsub$$
This is not hard to see.
First, the diffeomorphism, being generated by a hamiltonian vector,
is a canonical transformation and therefore preserves
the symplectic form
$$\L_{{}_{V_\alpha}} \omega = 0 \quad\Rightarrow\quad
g_t^* \omega = \omega . \eqn\invsymp$$
Moreover, any vector $v_t^{(i)}$ tangent to $L_t$ must be
the pushforward of some vector $v^{(i)}$ tangent to $L$, that is
$v_t^{(i)} = g_{t*} v^{(i)}$. Therefore
$$\omega (v_t^{(1)} , v_t^{(2)}) = \omega (g_{t*}v^{(1)},g_{t*}v^{(2)})
=g_t^* \omega (v^{(1)},v^{(2)}) = \omega (v^{(1)},v^{(2)}) = 0,
\eqn\sltisl$$
showing that indeed $L_t$ is lagrangian.

The first part of Eqn.\varclas\ implies that
$$ \bigl( S + \{ S , t\alpha \} \bigr) \big\vert_{p\in L}  =
S \big\vert_{g_t(p)\in L_t},\eqn\tcatgf$$
that is, the perturbed action on the left hand side, evaluated at
a point $p$ in the lagrangian submanifold, equals the original action
evaluated at the point $g_t(p)$ in the new lagrangian submanifold.
Thus, at the classical level ($\hbar=0$), $S$ and
$S+\delta_{t\alpha}S$, give the same gauge fixed theory when one
uses different gauge fixing conditions.

For the full quantum theory we claim that for {\it any} operator $A$
(whether or not it is an observable)
the following result holds:
$$\int_L d\lambda (A + \{A ,t\alpha \} ) \, e^{S +\delta_{t\alpha}S}
= \int_{L_t} d\lambda A \, e^S  +  \, \O(t^2).\eqn\bresult$$
If $A$ is an observable, the above equation does not give
a new result. Indeed \getinv ,
plus the independence of observables on the gauge fixing surface imply
\bresult\ for the case of observables.

The significance of \bresult\ is that off-shell quantities evaluated
with the modified action $S+\delta_{t\alpha}S$ are reproduced with
the original action by simply using a
{\it different gauge fixing surface}.
The two gauge fixing surfaces are related by the action of the
diffeomorphism arising from the function $t\alpha$. Note, however, that
as the action is modified, the operator whose expectation value is
being computed must also be modified. If we think of the supermanifold
$\M$ with its structure $(\omega , d\mu)$ fixed, an off-shell theory
is defined by the pair $(S,L)$, that is, an action and a lagrangian
submanifold.
The above equation shows that there is a map between the operators
of the {\it off-shell} theories $(S,L_t)$
and $(S + \delta_{t\alpha}S,L)$
such that the expectation values are identical.

In order to establish \bresult\ we first derive a formula relating
the measures of integration over the lagrangian submanifolds $L$
and $L_t$. We claim that
$$g_t^* \,d\lambda = (1+ t \Delta_\rho \alpha ) \, d\lambda + \O (t^2).
\eqn\clmeas$$
Consider the volume element $d\mu$. It follows from the
definition of Lie derivatives that
$$\eqalign{
g_t^* \, d\mu &= d\mu + t \L_{{}_{V_\alpha}}\, d\mu + \O (t^2) \cr
&= d\mu \, (1+ t\, \hbox{div}_\rho\, V_\alpha ) + \O (t^2) \cr
&= d\mu \, ( 1 + 2t\,\Delta_\rho \alpha ) + \O (t^2) ,\cr }
\eqn\relmestot$$
where use was made of the relation between the Lie derivative
of the volume element and the divergence operator, and of Eqn.\DELTA.
Consider now a set of basis vectors
$(e_1,\ldots,e_n,f^1,\ldots,f^n)$ for the tangent space
$T_{z\in L}\M $ such that $(e_1,\ldots,e_n)$ is a basis of
$T_{z\in L}L$ and the condition $\omega(e_i,f^j)=\delta_i^j$
is satisfied.
We then have that $\omega (g_{t*}e_i,g_{t*}f^j) = \delta_i^j$
(using Eqn.\invsymp ). Therefore
$$\eqalign{
g_t^*\, d\lambda (e_1,\cdots ,e_n) &=d\lambda \,(g_{t*}e_1,\cdots ,
g_{t*}e_n) \cr
&= \bigl(\, d\mu (g_{t*}e_1,\cdots ,g_{t*}e_n ; g_{t*}f^1, \cdots ,
g_{t*}f^n )\, \bigr)^{1/2}\cr
&= \bigl(\, g_t^* d\mu \,
(e_1,\cdots ,e_n;f^1,\cdots ,f^n)\, \bigr)^{1/2}\cr
&= \bigl( (1+2t\Delta_\rho \alpha ) \cdot
d\mu\, (e_1,\cdots ,e_n;f^1,\cdots ,f^n)\bigr)^{1/2} + \O (t^2)\cr
&= (1+t\Delta_\rho \alpha ) \cdot  d\lambda\,
(e_1,\cdots ,e_n) + \O (t^2), \cr }\eqn\allass$$
where use was made of Eqn.\LAMBDA. This establishes the validity
of Eqn.\clmeas.

We can now establish easily the desired relation (Eqn.\bresult ).
We have that
$$\int_{L_t} d\lambda\, A \, e^S = \int_L g_t^* (d\lambda \, A \,e^S )
= \int_L g_t^* (d\lambda )\, g_t^* ( A ) \, g_t^* (e^S ).
\eqn\gettopr$$
Making use of \pullscal\ and \clmeas\ we have that
$$\int_{L_t} d\lambda\, A \, e^S =
\int_L \, d\lambda  (1 + t \Delta_\rho \alpha ) (A + \{ A , t\alpha \})
e^{S+t\{ S, \alpha \} } , \eqn\gettopri$$
and Eqn.\bresult\ now follows simply by exponentiation of the
measure factor. This concludes our derivation.

When we use Darboux coordinates
$(\phi^i,\phi_i^*)$ and
the lagrangian submanifold $L$ specified by the gauge fermion
$\Upsilon(\phi)$ (as in Eqn.\eeffxd ),
the deformed lagrangian submanifold $L_t$ is defined by the equation
$$
\phi^*_i + t\Diff{_l \alpha(\phi,\phi^*)}{\phi^i}
=\left(\Diff{_l \Upsilon}{\phi^i}\right)
(\phi-t\diff{_l\alpha}{\phi^*})
=\Diff{_l \Upsilon(\phi)}{\phi^i}
-t\Diff{_l\alpha(\phi,\phi^*)}{\phi^*_j}
\DDiff{_l\Upsilon(\phi)}{\phi^j}{\phi^i} ,
\eqn\LTINDB
$$
which is obtained by using \inddif\ to relate the new variables to
the old ones. The last right hand side is obtained by expanding the
first right hand side.
Solving Eqn.\LTINDB\ for $\phi^*_i$ to first order in $t$,
we see that $L_t$ is given by
$\phi^*_i=\diff{_l\Upsilon_t(\phi)}{\phi^i}$
with the new gauge fermion $\Upsilon_t(\phi)$ given as
$$\Upsilon_t(\phi)=\Upsilon(\phi)
- t\alpha\, \bigl( \phi,\phi^*=\Diff{_l\Upsilon}{\phi}\bigr) .
\eqn\NEWUP$$
This result can also be derived from the variation of the
constraints fixing the lagrangian submanifold (see Eqn.\invol) under
the canonical transformation. Indeed, the new lagrangian submanifold
is defined by $G_i' = G_i + t \{ G_i , \alpha \} = 0$.

\section{Finite Symmetry Transformations}

The analysis of the previous subsection showed that the origin
of the $\de$ transformation was an infinitesimal canonical
transformation of field variables. This suggests that the
finite version of these transformations must arise from
finite canonical transformations.

Let $g: \M \rightarrow \M$ be a diffeomorphism that is
canonical, that is, it satisfies $g^*\omega = \omega$.
Explicitly, we write $z^I \rightarrow g^I(z)$.
If $g$ takes the lagrangian submanifold $L$ into the lagrangian
submanifold $L^g$, we then have
$$\int_{L^g}\!d\lambda\,e^S = \int_L\!g^* (d\lambda ) g^* (e^S)
= \int_L\!g^* (d\lambda )\,e^{S\left(g(z)\right)}.\eqn\oops$$
Following the logic of the previous subsection, if we can express
$g^* (d\lambda )$ as a factor multiplying $d\lambda$, then
we can exponentiate this factor and obtain the transformation
law of the action. We must first consider $d\mu = \rho(z) \prod dz^I$,
for which we have
$$\eqalign{
g^* d\mu &= \rho (g(z))\,\hbox{sdet}
\Bigl( {\partial_lg^I\over\partial z^J} \Bigr) \prod dz^I \cr
&= {\rho (g(z))\over \rho (z)}\,\hbox{sdet}
\Bigl( {\partial_l g^I \over \partial z^J} \Bigr) d\mu ,\cr}
\eqn\trmeas$$
and therefore
$$g^* d\mu = \left(F_g(z)\right)^2 d\mu , \quad \hbox{with} \quad
(F_g(z))^2 \equiv {\rho (g(z))\over \rho (z)}\,\hbox{sdet}
\Bigl( {\partial_lg^I\over\partial z^J} \Bigr) .\eqn\rmeas$$
Since the transformation induced by $g$ is canonical, the
analysis in the first three lines of Eqn.\allass\ applies,
and we obtain
$$g^* d\lambda = F_g(z)\,d\lambda ,\eqn\destra$$
and therefore back in Eqn.\oops\ we get
$$\int_{L^g}\!d\lambda\,e^S =
\int_L\!d\lambda\,e^{S\left(g(z)\right) + F_g(z)}.\eqn\aswewant$$
If $S$ satisfies the master equation, the integral on the left
hand side is independent of the choice of the lagrangian
submanifold, and as a consequence the integral on the right
hand side is also independent of this choice. Therefore, the
object in the exponential must satisfy the master
equation. Thus, the finite symmetry transformation generalizing
$\de$ is given by
$$S(z) \rightarrow S^g(z) \equiv S(g(z)) + \ln F_g(z) .\eqn\FINITEDE$$
For an infinitesimal canonical transformation
$g^I(z)=z^I + \{z^I,\e\}$, the transformation \FINITEDE\ can be
checked to reduce to $\de S$ to $\O (\e )$. In Appendix C, we show
explicitly that the actional $\A(S)$ of Eqn.\ameee\ is invariant
under these transformations: $\A(S^g)=\A(S)$.
Finally, the master equation $M(S)=0$ transforms nicely; one finds
$M(S^g) = g^*M(S)$. This is the generalization of Eqn.\onever\
for finite transformations.

%%%%%%%%%%%%%%%%%%%%%%%%%%%%%%%%%%%%%%%%%%%%%%%%%%%%%%%%%%%%%%%%%%%%%

\chapter{Changing the Decomposition of Moduli Space}

It is not difficult to obtain examples of string field theories that
make use of different decompositions of moduli space. One particularly
simple example is string field theories built with stubs of
different lengths (the definition of stubs is reviewed in \S4.6).
For two different values $l$ and $l'$ of this length, the subspaces
$\V_{g,n}(l)$ and $\V_{g,n}(l')$ of $\widehat\P_{g,n}$ defining the
string field vertices are different.
The resulting theories, however, have the same physical
content. More generally, we expect that any two string field
theories based on different decompositions should be physically
equivalent. Stub length is just one of an infinite number of parameters
that parameterize a generic deformation of the subspaces $\V_{g,n}$.

In this section we will prove that any deformation of the
$\V_{g,n}$'s used to build a string field action $S$ induces a change
of the form studied in \S3, namely,
$\delta_\e S = \Delta \e + \{ S, \e \}$ for some parameter $\e$ that
we will find. Since this change in the action is infinitesimal, we
will consider the infinitesimal problem. More precisely, assume we
have a one parameter family of subspaces $\V_{g,n}(u)$
with $u \in [0,1]$ giving us a family of
decompositions of moduli space, and as a consequence a one parameter
family of string field actions $S(u)$.
We will assume that the subspaces $\V(u)$ satisfy the
symmetry conditions demanding that the assignment of local coordinates
around the punctures be independent of the labels of the punctures.
Consistency demands that they must satisfy
the geometrical consistency conditions [\sonodazwiebach,\zwiebachlong]:
$$\partial \V_{g,n}(u) = -\partial_p R_1 \bigl( \V (u) \bigr) ,
\eqn\gccon$$
where the left hand side denotes the boundary
of the subspace $\V_{g,n}(u)$, and the right hand side denotes the
propagator boundary (sewing with sewing parameter $t$ satisfying
$|t| = 1$) of the subspace $R_1$ consisting of surfaces build out
of lower dimensionality subsets $\V$'s and a single sewing operation.

This section is organized as follows. In \S4.1 we review and
elaborate on some necessary tools from the operator formalism.
We then introduce in \S4.2 two vector fields $\widehat V$ and
$\widehat U$ that arise naturally in studying the deformation
of the subspaces $\V_{g,n}$. In \S4.3 we calculate the change
in $S$ induced by the deformation of the subspaces, and in \S4.4
we give the expression for the parameter $\e$ and
begin the verification that $\delta_\e S$ reproduces the
change of the action.
We explain how subspaces $\V_{g,n}$
that are not sections in $\widehat \P_{g,n}$ can yield a
consistent string field theory, and
discuss the algebra of $\de$ transformations (\S4.5). We
then study the deformation problem for the particular case of
stubs (\S4.6). Finally, in \S4.7 we complete our proof.

%%%%%%%%%%%%%%%%%%%%%%%%%%%%%%%%%%%%%%%%%%%%%%%%%%%

\section{Some Facts from the Operator Formalism}

We now review and elaborate somewhat on some of the tools
of the operator formalism relevant for string field theory
(for complete definitions see [\zwiebachlong]).
The results that will be given here are necessary for our
discussion in later sections. The basic objects are differential
forms in the tangent space $T_{\widehat\Sigma}\widehat\P_{g,n}$
based at the surface $\widehat\Sigma$. If we let $d_{g,n}$ denote
the real dimension of $\M_{g,n}$, then $\Omega^{(k)g,n}$ denotes
a $(d_{g,n}+k)$-form (for any $k\geq -d_{g,n}$) and is
labeled by $n$ arbitrary off-shell string fields
$${\Omega_{}^{}}^{(k)g,n}_{\Psi_1\cdots\Psi_n}(\widehat V_1,\cdots ,
\widehat V_{d_{g,n}+k} )
=N_{g,n} \bra{\Sigma}{\bf b}({\bf v}_1)\cdots
{\bf b}({\bf v}_{d_{g,n}+k})\ket{\Psi_1}\cdots\ket{\Psi_n},
\eqn\cdefform$$
with $N_{g,n} = (2\pi i)^{-(3g-3+n)}$ the normalization factor.
The Schiffer vector ${\bf v}_i= (v_i^{(1)},\cdots v_i^{(n)})$
creates the deformation specified by the tangent $\widehat V_i$,
and the antighost insertions are defined by
$${\bf b}({\bf v}) = \sum_{i=1}^n \biggl(
\oint b^{(i)}(z_i) v^{(i)}(z_i) {dz_i\over 2\pi i}
+\oint \overline b^{(i)}( \overline z_i)  \overline v^{(i)}
(\overline z_i) {d\overline z_i\over 2\pi i} \biggr).\eqn\kjhkjh$$
Similarly, given a Schiffer vector one defines the following
insertion of the stress tensor
$${\bf T}({\bf v}) = \sum_{i=1}^n \biggl(
\oint T^{(i)}(z_i) v^{(i)}(z_i) {dz_i\over 2\pi i}
+\oint \overline T^{(i)}( \overline z_i)  \overline v^{(i)}
(\overline z_i) {d\overline z_i\over 2\pi i} \biggr).\eqn\xdkjhkjh$$
The above forms satisfy the basic identity (Ref.[\zwiebachlong],
Eqn.(7.49))
$${\Omega_{}^{}}^{(k+1)g,n}_{(\sum Q)\Psi_1\cdots\Psi_n}
= (-)^{k+1}\,\hbox{d}{\Omega_{}^{}}^{(k)g,n}_{\Psi_1\cdots\Psi_n},
\eqn\qivbu$$
which says that the BRST operator $Q$ acts as an exterior derivative
on the extended moduli space $\widehat \P$.
Moreover, from the conventional definition of the contraction
operator $i_{{}_{\widehat U}}$
$$\bigl( {i_{}}_{\widehat U} \Omega \bigr)
(\widehat V_1 ,\cdots , \widehat V_k ) \equiv
\Omega (\widehat U , \widehat V_1 ,\cdots , \widehat V_k ),
\eqn\contraf$$
applicable to any form $\Omega$ (with $\widehat V_i$ arbitrary
vector fields) we readily find that for our special forms in
Eqn.\cdefform , one has
$$i_{{}_{\widehat U}}
{\Omega_{}^{}}^{(k+1)g,n}_{\Psi_1\cdots\Psi_n}
=(-)^k\, {\Omega_{}^{}}^{(k)g,n}_{{\bf b}({\bf u})\Psi_1\cdots\Psi_n},
\eqn\ivbu$$
where ${\bf u}$ is the Schiffer vector associated to the tangent
$\widehat U$.
The last identity we need involves the Lie derivative
$$\eqalign{
\L_{{}_{\widehat U}}\, {\Omega_{}^{}}^{(k)g,n}_{\Psi_1\cdots\Psi_n}
&\equiv \bigl( i_{{}_{\widehat U}} \hbox{d} +
\hbox{d} i_{{}_{\widehat U}} \bigr)\,
{\Omega_{}^{}}^{(k)g,n}_{\Psi_1\cdots\Psi_n} \cr
&= (-)^{k+1} \biggl( \, i_{{}_{\widehat U}}
{\Omega_{}^{}}^{(k+1)g,n}_{(\sum Q)\Psi_1\cdots\Psi_n} \, +\, \hbox{d}
{\Omega_{}^{}}^{(k-1)g,n}_{{\bf b}({\bf u})\Psi_1\cdots\Psi_n}\biggr)\cr
&= - \,{\Omega_{}^{}}^{(k)g,n}_{{\bf b}({\bf u})(\sum Q)
\Psi_1\cdots\Psi_n}
\,-\, {\Omega_{}^{}}^{(k)g,n}_{(\sum Q ){\bf b}({\bf u})
\Psi_1\cdots\Psi_n}\cr
&= -\, {\Omega_{}^{}}^{(k)g,n}_{\{ {\bf b}({\bf u}),\sum Q\}
\Psi_1\cdots\Psi_n},
\cr}\eqn\shlvbu$$
where use was made of Eqns.\qivbu\ and \ivbu.
Finally, since the anticommutator of the BRST operator and the
antighost field is the stress tensor, we find
$$ \L_{{}_{\widehat U}}\, {\Omega_{}^{}}^{(k)g,n}_{\Psi_1\cdots\Psi_n}
= -\, {\Omega_{}^{}}^{(k)g,n}_{{\bf T}({\bf u})
\Psi_1\cdots\Psi_n}.\eqn\lvbu$$
One final comment concerns notation. When the off-shell states that
label the forms are all the same we will define
$${\Omega_{}^{}}_{\underbrace{\scriptstyle \Psi\cdots\Psi}
_{\scriptstyle n}} \equiv {\Omega_{}}_{\Psi^n}.\eqn\notast$$

%%%%%%%%%%%%%%%%%%%%%%%%%%%%%%%%%%%%%%%%%%%%%%%%%%%

\section{The Generating Vectors $\widehat V$ and $\widehat U$}

\FIG\famcell{We show here a family of sections $\Gamma_{g,n}(u)$
that give rise to cell decompositions of moduli space. The base space
is $\M_{g,n}$ and the total space is $\widehat\P_{g,n}$.
We show the sections at $u_0$ and at $u_0+du$ and we indicate
the subsets $\V_{g,n}(u_0)$ and $\V_{g,n}(u_0 + du)$.
The parametrization by $u$ of the family of sections induces
a parametrization by $u$ on the fibers. (a) The vector $\widehat V$
is the vertical vector along the fibers. (b) The vector $\widehat U$
is a vector field that generates a diffeomorphism mapping the
subspaces $\V_{g,n}$ into each other.}

In this subsection we want to describe some of the geometrical
structure available; in particular, we will introduce two vector
fields, called $\widehat V$ and $\widehat U$, that are relevant
to the deformation of the subspaces $\V_{g,n}$.

The situation we have in mind is illustrated in Fig.\famcell ,
where the family of sections $\Gamma_{g,n}(u)$ is shown. The base space
is moduli space $\M_{g,n}$ and the total space is $\widehat\P_{g,n}$.
The subsets $\V_{g,n} (u)$ are indicated schematically, in particular
$\V_{g,n}(u_0)$ and $\V_{g,n}(u_0+du)$.
The fibers, representing surfaces which have the same conformal
structure but different coordinate systems at the punctures,
are also parametrized by $u$.
For any point $p \in \M_{g,n}$, the fiber over $p$ is a curve
denoted as $f_p(u)$, where $f_p : [0,1 ]\rightarrow \widehat \P_{g,n}$,
and $f_p(u) \in \Gamma_{g,n}(u)$.
We define the vertical vector field $\widehat V$ to be the tangent
vector to the fibers; more precisely
$\widehat V=f_{p*}(\partial/\partial u)$.
The vector field $\widehat V$ is the vector which generates the
diffeomorphisms moving the sections; we let
$f^{\widehat V}_t$ denote the diffeomorphism which moves any point
in $\widehat \P$ a parameter distance $t$ along the fiber. Thus
$f^{\widehat V}_t (\Gamma_{g,n}(u)) = \Gamma_{g,n}(u+t)$.
(We will omit the subscripts $\{ g,n\}$ when there is no room for
confusion.)

Consider now the neighboring sections $\Gamma (u_0)$ and
$\Gamma (u_0+du)$, and the relevant subspaces
$\V (u_0)$ and $\V (u_0+du)$.
Define now the (oriented) subspace of $\Gamma (u_0+du)$
$$\delta_{u_0}\V_{g,n} \equiv \V_{g,n}(u_0 + du) - f^{\widehat V}_{du}
\bigl( \V_{g,n}(u_0) \bigr) . \eqn\diffspace$$
This space is the difference between $\V (u_0+du)$ and the image
of $\V (u_0)$ in $\Gamma (u_0+du)$. It includes with a plus sign
the subspace of $\Gamma (u_0+du)$ corresponding to the surfaces that,
regardless of the local coordinates, are contained in $\V (u_0+du)$
but not in $\V (u_0)$, and with minus sign the subspace corresponding
to the surfaces that, regardless of the local coordinates, are in
$\V (u_0)$ but are not in $\V (u_0+du)$. This subspace will be
relevant for us in the next subsection.

Let us now define a vector field $\widehat U$. This vector
field will not be uniquely determined
at this stage. Our assumption that the subsets
$\V_{g,n}(u)$, as we change $u$, are smoothly related will be taken
to mean that there is a family of diffeomorphisms connecting them.
The vector field $\widehat U$ generates such diffeomorphisms; that is,
it pushes the subspaces $\V_{g,n}(u)$ precisely into
each other. This is illustrated in Fig.\famcell (b), where
the curve $h_p(u)$ is the trajectory followed by the point
$p$ representing a surface in $\V_{g,n}(u_0)$.
More precisely, the vector field $\widehat U$ is defined to be the
tangent vector $\widehat U=h_{p*}(\partial/\partial u)$.
The corresponding diffeomorphisms will be denoted by $f^{\widehat U}_s$
and they map
$$\eqalign{
f^{\widehat U}_s: &\quad \V_{g,n}(u_0) \, \rightarrow\,
\V_{g,n}(u_0+s);\cr
f^{\widehat U}_s: &\quad\partial\V_{g,n}(u_0)\, \rightarrow\,
\partial\V_{g,n}(u_0+s);\cr}
\eqn\oblmap$$
We will discuss later how to make a convenient choice of vector
field $\widehat U$ by requiring compatibility with sewing.

%%%%%%%%%%%%%%%%%%%%%%%%%%%%%%%%%%%%%%%%%%%%%%%%%%%

\section{Change in $S$ due to a Deformation of the $\V_{g,n}$'s.}

We are now ready to calculate the variation of the action.
An infinitesimal change of $u$ induces the change
$$S (u_0 + du ) = S(u_0) + du\cdot {dS\over du}\bigg|_{u_0} + \O (du^2),
\eqn\changeact$$
and we want to show that
$$\delta S(u_0) \equiv du\cdot {dS\over du}\bigg|_{u_0} =
\hbar\Delta \e + \{ S(u_0), \e \}, \eqn\wtshow$$
for some $\e$ of the form $\e = du\!\cdot\!e(u_0)$.
This will establish that
infinitesimal changes of the cell decomposition are generated by
the transformations discussed in the previous section.
Such transformations could, in principle, be integrated to prove that
$S(u=0)$ and $S(u=1)$ are related by a large transformation, but we will
not discuss this explicitly. The purpose of the present subsection is
simply to evaluate the left hand side of \wtshow. This is what
we do next.

The closed string action is given by
$$S(u,\Psi ) = {1\over 2}\, \langle \Psi , Q \Psi \rangle
+ \sum_{g,n} \hbar^g \kappa^{n+2g-2}
\, S_g^n (u,\Psi ),\eqn\yesxfxf$$
where the sum extends over $n\geq 3$ for $g=0$, and over
$n\geq 1$ for $g\geq 1$. Here $S_g^n$ is defined by
the following expression [\zwiebachlong]
$$S_g^n (u, \Psi ) = {1\over n!} \hskip-6pt\int_{\V_{g,n}(u)}
\hskip-6pt \Omega^{(0)g,n}_{\Psi^n} .\eqn\termact$$
The change we are considering does not affect the kinetic
term $S_0^2$ and therefore the variation of the action will be given by
$$\delta S(u,\Psi ) = \sum_{g,n} \hbar^g \kappa^{n+2g-2}
\, \delta S_g^n (u,\Psi ).\eqn\xfxf$$
Making use of Eqn.\termact\ we find that the variation of $S_g^n$
is given by
$$\eqalign{
n!\, [S_g^n (u_0+du,\Psi )-S_g^n(u_0,\Psi )] &=
\int_{\V_{g,n}(u_0+du)}\hskip-6pt\Omega^{(0)g,n}_{\Psi^n}
-\int_{\V_{g,n}(u_0)} \Omega^{(0)g,n}_{\Psi^n} \cr
&= \int_{\V_{g,n}(u_0)}\hskip-6pt\bigl(
f^{\widehat U\, * }_{du} \Omega^{(0)g,n}_{\Psi^n}
-\Omega^{(0)g,n}_{\Psi^n}\bigr), \cr}\eqn\yuiu$$
where $ f^{\widehat U\, * }_{du} \Omega$ denotes the pullback form
using the diffeomorphism generated by the vector field $\widehat U$,
mapping the subspaces precisely into each other.
We note that the difference of forms on the (last) right hand
side relates to the Lie derivative
of the form $\Omega$ along the vector field $\widehat U$. Indeed,
taking the limit $du \rightarrow 0$ we find
$${dS_g^n \over du}\bigg|_{u_0} = {1\over n!}\hskip-6pt
\int_{\V_{g,n}(u_0)} \L_{\widehat U}\,
\Omega^{(0)g,n}_{\Psi^n}\, .\eqn\ssyuiu$$
This is the total change in $S_g^n$. The case of $S_0^3$ is special
because $\V_{0,3}(u)$ is simply a point (for each $u$). In this case
the variation is simply given by the integrand in the above expression.

We could have made use of the diffeomorphism $f^{\widehat V}_{du}$
generated by the vertical vector field $\widehat V$ in order to
compute the change of the action. In this case we would have
$$\int_{\V_{g,n}(u_0+du)}\hskip-6pt\Omega^{(0)g,n}_{\Psi^n}
= \int_{\delta_{u_0}\V_{g,n}} \Omega^{(0)g,n}_{\Psi^n}
+ \int_{\V_{g,n}(u_0)}\hskip-6pt
f^{\widehat V\, * }_{du} \Omega^{(0)g,n}_{\Psi^n}, \eqn\dfyuiu$$
where use was made of Eqn.\diffspace. We then find that the
change in the action can be written alternatively as
$${dS_g^n \over du}\bigg|_{u_0} = {1\over n!}\hskip-6pt
\int_{\V_{g,n}(u_0)} \L_{\widehat V}\,
\Omega^{(0)g,n}_{\Psi^n}\,
+{1\over n!} \lim_{du\rightarrow 0}\, {1\over du}\cdot\hskip-6pt
\int_{\delta_{u_0}\V_{g,n}} \Omega^{(0)g,n}_{\Psi^n}.\eqn\uuiu$$

Having these two alternative expressions for the variation of
the action will be helpful to understand some of the relevant
issues.

%%%%%%%%%%%%%%%%%%%%%%%%%%%%%%%%%%%%%%%%%%%%%%%%%%%

\section{Construction of the Symmetry Generator}

We have computed in the previous subsection the
derivative $dS/du$ of the action as we change the sections
in the spaces $\widehat\P_{g,n}$ (or equivalently,
the relevant subspaces $\V_{g,n}$).
The purpose of the present subsection is to find the
generator of the symmetry transformation that reproduces this
change. From \wtshow\ we must find a parameter $e(u_0)$ such that
$${dS\over du}\bigg|_{u_0} = \hbar\Delta e(u_0) +\{ S(u_0) , e(u_0)\}.
\eqn\swtshow$$
We claim that the answer is very simple. One must have
$$ e(u_0) = \sum_{g,n} \hbar^g \kappa^{n+2g-2}
\,  e_g^n (u_0,\Psi ),\eqn\xxfxf$$
where, as before, the sum extends over $n\geq 3$ for $g=0$ and
$n\geq 1$ for $g\geq1$, and where
$e_g^n$ carries the information about the change in subspace
at genus $g$ and $n$ punctures. Since both the delta operator
and the antibracket have ghost number $+1$, Eqn.\swtshow\ requires
that $e_g^n$ must carry ghost number $-1$, and therefore it must
include an extra antighost insertion, or equivalently, it
must be related to the contraction $i_X\Omega^{(+1)}$ with some
vector $X$. We have found in the previous subsection two vectors
$\widehat U$ and $\widehat V$. Thus we are led to consider the
possibility that
$$e_g^n(u_0,\Psi) = - {1\over n!} \int_{\V_{g,n}(u_0)}
\hskip-6pt i_{{}_{\widehat U}}
{\Omega_{}^{}}^{(+1)g,n}_{\Psi^n} ,\eqn\answeryes$$
or that instead
$$e_g^n(u_0,\Psi) = - {1\over n!} \int_{\V_{g,n}(u_0)}
\hskip-6pt i_{{}_{\widehat V}}
{\Omega_{}^{}}^{(+1)g,n}_{\Psi^n}.\eqn\aweryes$$
Both candidates are quite similar; in both cases one takes
the form $\Omega^{(+1)}$, of degree one higher than that suitable
for integration over the section, and contracts it with a  vector field
to get a form that we can integrate over the original subspace
$\V_{g,n}(u_0)$.
Actually, both expressions are {\it the same} to order $\e$ and are
therefore completely equivalent for our purposes. This happens
because the vector $\widehat U$ can be decomposed into a component
along the fibers, which coincides with $\widehat V$ to order $\e$, and
a component $\widehat W$ tangent to the space $\V_{g,n}(u_0)$. This
second component cannot contribute to the integration because of the
antisymmetry of the form $\Omega$, which already has as input a basis
of vectors for the tangent space of $\V_{g,n}$ over which it is being
integrated.

Making use of Eqn.\ivbu\ we can rewrite Eqn.\answeryes\ as
$$e_g^n(u_0,\Psi) = - {1\over n!} \int_{\V_{g,n}(u_0)}
\hskip-6pt {\Omega_{}^{}}^{(0)g,n}_{{\bf b}({\bf u})\Psi^n} .
\eqn\weryes$$
We can now begin our proof that $e(u_0)$ defined above
does indeed generate the correct change in the action.
Making use of Eqns.\yesxfxf, \swtshow\ and \xxfxf\ we have to show that
$${dS_g^n\over du}\bigg|_{u_0} =\Delta e_{g-1}^{n+2}(u_0)
+\sum\limits_{{g_1+g_2=g}\atop {n_1+n_2 = n+2}}
\{ S_{g_1}^{n_1}(u_0), e_{g_2}^{n_2}(u_0) \}.\eqn\tscvhow$$
Our aim is now to evaluate the right hand of this equation
and to prove that it coincides with the left hand side given
earlier in \ssyuiu. We will be using explicitly the vector
$\widehat U$ in our computation.

Let us start with the $\{ S , e \}$ type term, and in particular
we single out the contribution from $S_0^2$. A computation
(analogous to that in \S4 of Ref.[\zwiebachlong]) gives us
$$\eqalign{
\bigl\lbrace S_0^2 , e_g^n \bigr\rbrace &=
-{1\over n!}\int_{\V_{g,n}(u_0)}
{\Omega_{}^{}}^{(0)g,n}_{{\bf b}({\bf u})(\sum Q)\Psi^n} \cr
&= -{1\over n!} \int_{\V_{g,n}(u_0)}
\Omega^{(0)g,n}_{{\bf T}({\bf u})\Psi^n}
+{1\over n!}\int_{\partial\V_{g,n}(u_0)}
\Omega^{(-1)g,n}_{{\bf b}({\bf u})\Psi^n}\,\,  ,\cr}\eqn\tnks$$
where in the last step we made use of \qivbu. Eqns.\ivbu\ and
\lvbu\ then give
$$\bigl\lbrace S_0^2 , e_g^n \bigr\rbrace
={1\over n!}\int_{\V_{g,n}(u_0)}
\L_{{}_{\widehat U}} \Omega^{(0)g,n}_{\Psi^n}
-{1\over n!}\int_{\partial\V_{g,n}(u_0)}
i_{{}_{\widehat U}}\Omega^{(0)g,n}_{\Psi^n}.
\eqn\xstnks$$
Note that the first term on the right hand side is precisely
the Lie derivative term appearing on the right hand side of \ssyuiu.
Let us now consider the contributions to $\{ S , e\}$
for $S$ different from the kinetic term. This time a calculation gives
$$\eqalign{
\bigl\lbrace S_{g_1}^{n_1} , e_{g_2}^{n_2} \bigr\rbrace
&= -{1\over (n_1-1)!} {1\over (n_2-1)!}
\cdot \hskip-6pt \int_{\V_{g_2,n_2}(u_0)} \hskip-8pt
i_{{}_{\widehat U}}\Omega^{(+1)g_2,n_2}_{\Psi^{n_2-1}
[\Psi^{n_1-1}]_{g_1}} \cr
&=  -{1\over (n_1-1)!} {1\over (n_2-1)!}
\cdot{\sum_s}'(-)^{\Phi_s}\hskip-6pt
\int_{\V_{g_2,n_2}(u_0)} \hskip-8pt
i_{{}_{\widehat U}}{\Omega_{}^{}}^{(+1)g_2,n_2}_{\Psi^{n_2-1}
\widetilde\Phi_s}
\cdot \hskip-6pt\int_{\V_{g_1,n_1}(u_0)} \hskip-8pt
{\Omega_{}^{}}^{(0)g_1,n_1}_{\Phi_s\Psi^{n_1-1}}\,\, ,\cr}
\eqn\theterm$$
where in the last step we used the definition of the
string product (Ref.[\zwiebachlong], Eqn.(7.100)),
and the primed summation $\sum'$ implies the summation over the
states annihilated by $L_0^-$.
Finally for the term involving the $\Delta$ operator we get
$$\Delta\, e_{g-1}^{n+2} =-\,{1\over 2}\,{1\over n!}\,
{\sum_s}' (-)^{\Phi_s} \hskip-8pt
\int_{\V_{g-1,n+2}(u_0)} \hskip-8pt i_{{}_{\widehat U}}
{\Omega_{}^{}}^{(+1)g-1,n+2}_{\Psi^n \Phi_s \widetilde\Phi_s}.
\eqn\deltacont$$

We have now evaluated the basic ingredients appearing on the
right hand side of Eqn.\tscvhow. Comparing with Eqn.\ssyuiu\ we find
that the generator $e(u_0)$ will give the correct change
in the action if the following relation holds:
$$\eqalign{0 &=\quad {1\over n!}\hskip-6pt\int_{\partial\V_{g,n}(u_0)}
\hskip-6pt i_{{}_{\widehat U}}\Omega^{(0)g,n}_{\Psi^n}\cr
&\quad +\sum\limits_{{g_1+g_2=g}\atop {n_1+n_2 = n+2}}
{1\over (n_1-1)!} {1\over (n_2-1)!}
\cdot{\sum_s}'(-)^{\Phi_s}\hskip-6pt
\int_{\V_{g_2,n_2}(u_0)} \hskip-8pt
i_{{}_{\widehat U}}{\Omega_{}^{}}^{(+1)g_2,n_2}_{\Psi^{n_2-1}
\widetilde\Phi_s}
\cdot \hskip-6pt\int_{\V_{g_1,n_1}(u_0)} \hskip-8pt
{\Omega_{}^{}}^{(0)g_1,n_1}_{\Phi_s\Psi^{n_1-1}} \cr
&\quad + {1\over 2}\,{1\over n!}\,
{\sum_s}' (-)^{\Phi_s} \hskip-10pt
\int_{\V_{g-1,n+2}(u_0)} \hskip-8pt i_{{}_{\widehat U}}
{\Omega_{}^{}}^{(+1)g-1,n+2}_{\Psi^n \Phi_s \widetilde\Phi_s}.\cr}
\eqn\lotsh$$
It must be noted that in the second and third lines of \lotsh\ we
can replace the vector $\widehat U$ by the vertical
vector $\widehat V$, since the integrals already extend over
the full subspaces. This {\it cannot} be done in the first line,
since the integral extends only over the boundary of $\V_{g,n}$
and in the decomposition $\widehat U =\widehat V + \widehat W$
the vector $\widehat W$ need not be tangent to $\partial \V_{g,n}$.

If we had used the expression \aweryes\ for $e_g^n$ based on the vector
$\widehat V$, there would have been only one change in the above
derivation. Making use of Eqn.\uuiu\ one finds that the
first term in the above equation would have been replaced by
$${1\over n!}\hskip-6pt\int_{\partial\V_{g,n}(u_0)}
\hskip-6pt i_{{}_{\widehat V}}\Omega^{(0)g,n}_{\Psi^n}
+{1\over n!} \lim_{du\rightarrow 0}\, {1\over du}\cdot\hskip-6pt
\int_{\delta_{u_0}\V_{g,n}} \Omega^{(0)g,n}_{\Psi^n}.\eqn\xuiu$$
The vector $\widehat W du$ that we have been discussing is, to order
$\e$, the vector that moves the image (under the vertical map)
of the boundary of $\V_{g,n}(u_0)$ in the section $\Gamma (u_0+ du)$
to the boundary of $\V_{g,n}(u_0+du)$ in the same section. This
vector therefore maps out the space $\delta_{u_0}\V_{g,n}$ introduced
before. This fact allows us to rewrite
the second term in the above expression as
$${1\over n!} \lim_{du\rightarrow 0}\, {1\over du}\cdot\hskip-6pt
\int_{\partial\V_{g,n}} i_{{}_{\widehat W du}}
\Omega^{(0)g,n}_{\Psi^n}
= {1\over n!}\cdot\hskip-6pt
\int_{\partial\V_{g,n}} i_{{}_{\widehat W}}
\Omega^{(0)g,n}_{\Psi^n},\eqn\xxuuiu$$
combining this with the first term in \xuiu\ and making use of
$i_{{}_{\widehat V}} + i_{{}_{\widehat W}} = i_{{}_{\widehat U}}$
we obtain, as expected, the same expression for the first term
appearing in Eqn.\lotsh.

In the next two subsections we will develop intuitive understanding
of the expressions given in this subsection, providing partial
confirmation of the correctness of our ansatz for $e_g^n$.
In the last subsection we give a complete, though slightly more
abstract, proof of the result.

%%%%%%%%%%%%%%%%%%%%%%%%%%%%%%%%%%%%%%%%%%%%%%%%%%%

\section{Why Sections are not Absolutely Necessary}

\FIG\nosect{(a) On the space $\widehat\P_{g,n}$ over
$\M_{g,n}$ we show a subspace $\V_{g,n}$ which is a section over
$\M_{g,n}$, and a subspace $\V'_{g,n}$ which is not a section because
the vertical fibers intersect it more than once. The two subspaces
coincide at their boundaries and together they bound a region $R$.
(b) This time we show two sections, one at $u_0$ and the other
at $u_0+du$. The subspace $\V_{g,n}(u_0)$ extends from $A$ to $B$
and the subspace $\V_{g,n}(u_0+du)$ extends from $C$ to $D$.
We denote by ${\cal S}$ the `strip' joining the two subspaces from
their boundaries. The strip plus the subspace at $u_0+du$ form a
subspace which is not a section over $\M_{g,n}$.}

To date, string field theory has been constructed using subspaces
of sections on the bundle $\widehat \P_{g,n}$ over moduli space
$\M_{g,n}$. This means that we pick a subset of surfaces of $\M_{g,n}$
and for each such surface we give a {\it unique} surface in the
space $\widehat \P_{g,n}$, corresponding to choosing local coordinates
around the punctures. This corresponds to choosing a section, that
is a well defined map $\pi^{-1}: \M_{g,n}\rightarrow \widehat\P_{g,n}$.
In a more general situation $\V_{g,n}$ is simply a subspace
of $\widehat\P_{g,n}$, where it can happen that the projection $\pi$
may take more than one point in the subspace to the same point in
$\M_{g,n}$, as illustrated in Fig.\nosect (a). Such generalized
subspaces $\V_{g,n}$ will still lead to an action satisfying the BV
master equation, if the geometrical recursion relations, relating the
boundary of subspaces to sewing operations with lower dimensional
subspaces, still hold. For physical states the value of such a vertex
will be the same as that from a vertex defined by a section with the
same boundary. This is easily verified as follows. Consider a subspace
$\V'$ which is not a section, and the related (section) subspace $\V$
with coincident boundaries. Let $R$ denote the region bound by
$\V$ and $\V'$, that is, $\partial R = \V - \V'$. Since physical
states are annihilated by $Q$ we then have
$$0= \int_R \Omega^{(+1)g,n}_{(\sum Q)\Psi\cdots\Psi} =
-\int_R d\Omega^{(0)g,n}_{\Psi\cdots\Psi} =
-\int_{\V} \Omega^{(0)g,n}_{\Psi\cdots\Psi} +
\int_{\V'} \Omega^{(0)g,n}_{\Psi\cdots\Psi} ,\eqn\eeqph$$
establishing the desired equality.

We will now see that particular choices of $e_g^n$'s turn a string
field theory based on sections into one which does not use sections.
Since for any choice of $e$'s we {\it must} get a consistent string
field theory we simply have to study the effect of the change.

Suppose we take $e_{g_0}^{n_0} \not=0$ and all others are set to
zero. We also assume that this $e_{g_0}^{n_0}$ is of the form
given in Eqn.\answeryes\ for some vector field $\widehat U$.
The symmetry transformation will change $S_{g_0}^{n_0}$,
terms with a larger number of string fields for the same genus,
and terms of higher genus (see Eqn.\tscvhow ). Let us see what
happens with $S_{g_0}^{n_0}$. Its variation is controlled by the
bracket of $e_{g_0}^{n_0}$ with $S_0^2$. Thus making use of
Eqn.\xstnks\ we find
$$\eqalign{
{S'}_{g_0}^{n_0} &=S_{g_0}^{n_0}
+du\bigl\lbrace S_0^2,e_{g_0}^{n_0}\bigr\rbrace\cr
&=S_{g_0}^{n_0}+ {1\over n!}\hskip-6pt\int_{\V_{g,n}(u_0)}\hskip-8pt
\L_{{}_{\widehat U du}} \Omega^{(0)g,n}_{\Psi^n}
+{1\over n!}\hskip-6pt\int_{\partial\V_{g,n}(u_0)}
\hskip-6pt i_{{}_{-\widehat U du}}\Omega^{(0)g,n}_{\Psi^n}.\cr}
\eqn\nks$$
We can now combine the first two terms in the last expression to find
$${S'}_{g_0}^{n_0} = {1\over n!}
\hskip-6pt\int_{\V_{g,n}(u_0+du)}
\Omega^{(0)g,n}_{\Psi^n} + {1\over n!}
\hskip-6pt\int_{\partial\V_{g,n}(u_0)}\hskip-8pt
i_{{}_{-\widehat U du}}\Omega^{(0)g,n}_{\Psi^n} .\eqn\mnks$$
This expression has a simple meaning. The second term is nothing
else than the integral of the form $\Omega^{(0)}$ over the strip
${\cal S}$ shown in Fig.\nosect (b). This is the strip defined by the
vector $\widehat U$ over $\partial\V_{g,n}$. Therefore,
$${S'}_{g_0}^{n_0} = {1\over n!}
\hskip-6pt\int_{\V'_{g,n}}
\Omega^{(0)g,n}_{\Psi^n},\eqn\mmks$$
where the new subspace $\V'_{g,n}$ is the subspace including
the subspace at $(u_0+du)$ plus the strip, making it have the
same boundary as the original subspace at $u_0$. The new subspace
does not, in general, correspond to a section in $\widehat\P$.
The new action thus works without using a section, and this
illustrates not only how the symmetry transformation produces
consistent modifications, but why sections are not absolutely
required to have a consistent theory. Full appreciation of this
fact may be important to obtain generalized formulations of
string field theory.

We have remarked in Eqn.\lalgeb\ that the Lie algebra of the
$\de$ transformations is the Lie algebra of the antibracket.
The $\de$ transformations with $\e$'s corresponding to a change
in section are a small subset of all possible $\de$ transformations.
If $\e_1$ and $\e_2$ are two parameters corresponding to changes
in sections, their antibracket $\{ \e_1 , \e_2 \}$ does not
seem to be, in general, a parameter associated to
another change in section. It is simply another parameter with
some complicated effect on the vertices, an effect that can
be far more involved than taking sections into subspaces that
are not sections. This is expected since successive transformations
do not arise naturally in our context; our deformations are
not defined for all possible choices of subspaces, thus after
deforming a bit with a first vector $\widehat U_1$ we may obtain a
subspace for which a second vector $\widehat U_2$ may not be defined.
Therefore, the parameters associated with changes of section
do not seem to define a Lie subalgebra of the antibracket.
It would be very interesting to find special parameters associated
with changes of section that form a subalgebra.

%%%%%%%%%%%%%%%%%%%%%%%%%%%%%%%%%%%%%%%%%%%%%%%%%%%

\section{The particular case of Stubs}

The most straightforward way of creating a one parameter family
of subspaces $\V_{g,n}$ satisfying the consistency conditions
is based on the variation of the stub length in covariant closed
string field theory [\sonodazwiebach , \brusteinalwis]. With a minimal
area metric, each surface in the
subspace $\V_{g,n}$ has a semiinfinite cylinder about every
puncture, and the corresponding local coordinate $z$ is
defined by taking the curve $|z|=1$ to be the geodesic circle
a distance $l$ down the cylinder (measured from the beginning of the
semiinfinite cylinder). This short cylinder of length $l$ is called
the stub ($l \geq \pi$). If the string vertices have stubs of length
$l$, the subspaces $\V_{g,n}$ are given by all the surfaces whose
metric of minimal area does not show any finite cylinder of length
greater than $2l$ [\zwiebachlong]. This means that the surface has no
propagator, or that no sewing operation is involved in its construction.
This result tells us explicitly how
the subspaces $\V_{g,n}$ vary as we vary the stub length.

Consider an infinitesimal increase of the stub length by an amount
$\e$. The new surfaces that must be included in each $\V_{g,n}$
are those that have at least one cylinder with length
$L$ in the interval $2l \leq L \leq 2l+2\e$, and all other cylinders
of length smaller than $2l+2\e$. We claim that to order $\e$
we need only consider the surfaces with {\it one propagator only}.
This is clear because when we have one propagator
($2l \leq L \leq 2l+2\e$) the domain representing the new surfaces
has a `volume' proportional to the product of the `volumes' of two
subspaces involved in the sewing (tree configuration);
each time we have another propagator, since its parameter space is
proportional to $\e$, the volume is reduced by this factor,
and therefore we get a higher order effect.

Let us now give the explicit expression for the changes. Let $z$ be
the original local coordinates. Increasing the stub length by $\e$
corresponds to defining a new coordinate $z' = z + \e z$, since the
circle $|z'|=1$ corresponds to $|z|=1-\e$, and is therefore a
retraction of the local coordinate. The associated Schiffer vector
defined from $z' = z + e v(z)$ is then $v(z) =z$.
This Schiffer vector, used for each puncture in a given surface,
implements the deformation associated to the vertical vector
$\widehat V$ defined earlier, since its effect is simply to change
the local coordinates at the punctures, without changing the moduli
of the surface. The antighost insertion for any vertex with $n$
punctures is then given by
$${\bf b} ({\bf v}) = \sum_i ( b_0^{(i)} + \overline b_0^{(i)})
= \sum_i b_0^{+(i)},\eqn\chstub$$
and the stress tensor insertion is given by
$${\bf T} ({\bf v}) = \sum_i ( L_0^{(i)} + \overline L_0^{(i)})
= \sum_i L_0^{+(i)}.\eqn\chsten$$
Equation \uuiu\ then reads
$${dS_g^n \over dl}\bigg|_{l_0} = -\,{1\over (n-1)!}\hskip-6pt
\int_{\V_{g,n}(l_0)}
\Omega^{(0)g,n}_{(L_0^+\Psi ) \Psi^{n-1}}\,
+{1\over n!} \lim_{du\rightarrow 0}\, {1\over du}\cdot\hskip-6pt
\int_{\delta_{l_0}\V_{g,n}}\hskip-6pt\Omega^{(0)g,n}_{\Psi^n},
\eqn\huiu$$
where $l_0$ denotes the original stub length, and the
second term represents the contribution due to the
extra surfaces that must be included in the new section.
Moreover, our ansatz for $e_g^n$ reduces to
$$e_g^n(l_0,\Psi) = -\, {1\over (n-1)!} \int_{\V_{g,n}(l_0)}
\hskip-6pt {\Omega_{}^{}}^{(0)g,n}_{(b_0^+\Psi )\Psi^{n-1}}.\eqn\insc$$

Let us now argue that Eqn.\lotsh\ must be satisfied in this case.
We will only pay attention to how the relevant surfaces appear
and not to the combinatorial factors, which are verified to work
explicitly in the next subsection.
As argued below Eqn.\lotsh\ (see Eqns.\xuiu\ and \xxuuiu )
the first term on the right hand side equals
$${1\over n!}\hskip-6pt\int_{\partial\V_{g,n}(l_0)}
\hskip-6pt \Omega^{(-1)g,n}_{{\bf b}({\bf v})\Psi^n}
+{1\over n!}\cdot\hskip-6pt
\int_{\partial\V_{g,n}(l_0)}\Omega^{(-1)g,n}_{{\bf b}({\bf w})\Psi^n},
\eqn\wwiu$$
where ${\bf w}$ is the Schiffer vector generating the deformation that
gives the new surfaces.  This Schiffer vector, as usual, is supported
on the external punctures.
Since every surface $\Sigma$ in $\partial \V_{g,n}$
is built by sewing, and the deformations we are considering are
precisely due to a change of sewing parameter, we can relate external
insertions to internal ones. Given that such deformations of the
surface can be obtained in either way we must have that
$$\bra{\Sigma}{\bf T}({\bf w}) =
\bra{\widehat\Sigma_1}\bra{\widehat\Sigma_2}R^\theta_{r_1r_2}\rangle
{\bf T}({\bf w}) =
\bra{\widehat\Sigma_1}\bra{\widehat\Sigma_2}
\bigl( L_0^+ \ket{R^\theta_{r_1r_2}} \bigr) ,
\eqn\sscss$$
where $\ket{R^\theta_{r_1r_2}}$ is the sewing ket. The same equation
must hold for antighosts, namely
$$\bra{\Sigma}{\bf b}({\bf w}) =
\bra{\widehat\Sigma_1}\bra{\widehat\Sigma_2} \bigl(
b_0^+ \ket{R^\theta_{r_1r_2}} \bigr) . \eqn\pppss$$
We also noted below Eqn.\lotsh\ that we can use the vertical vector
$\widehat V$ in the second and
third lines of that equation. Let us consider the second line, which
up to numerical coefficients and the sums can be written as
$$\eqalign{
(-)^{\Phi_s}\hskip-10pt
\int_{\V_{g_2,n_2}(l_0)} \hskip-8pt
{\Omega_{}^{}}^{(0)g_2,n_2}_{
{\bf b}({\bf v})\Psi^{n_2-1}\widetilde\Phi_s}
\cdot \hskip-6pt\int_{\V_{g_1,n_1}(l_0)} \hskip-8pt
{\Omega_{}^{}}^{(0)g_1,n_1}_{\Phi_s\Psi^{n_1-1}}
&=(-)^{\Phi_s}\hskip-10pt
\int_{\V_{g_2,n_2}(l_0)} \hskip-8pt
{\Omega_{}^{}}^{(0)g_2,n_2}_{(\sum b_0^+\Psi^{n_2-1})\widetilde\Phi_s}
\cdot \hskip-6pt\int_{\V_{g_1,n_1}(l_0)} \hskip-8pt
{\Omega_{}^{}}^{(0)g_1,n_1}_{\Phi_s\Psi^{n_1-1}} \cr
&+(-)^{\Phi_s}\hskip-10pt
\int_{\V_{g_2,n_2}(l_0)} \hskip-8pt
{\Omega_{}^{}}^{(0)g_2,n_2}_{\Psi^{n_2-1}(b_0^+\widetilde\Phi_s)}
\cdot \hskip-6pt\int_{\V_{g_1,n_1}(l_0)} \hskip-8pt
{\Omega_{}^{}}^{(0)g_1,n_1}_{\Phi_s\Psi^{n_1-1}}. \cr}\eqn\newdec$$
The first term on the right hand side represents sewing of
surfaces and integration over the sum of direct products of subspaces
and twist angle. This total space is simply $\partial\V_{g,n}$.
Note that the antighost insertions are acting only on the external
punctures (by symmetrization using the implicit sums, they also
act on the external punctures of the surfaces in $\V_{g_1,n_2}$).
But the antighost insertions for a uniform change in stub length are
always $b_0^+$ regardless of genus or the number of punctures,
so this term is exactly of the same as the first term \wwiu.
The second term shows the insertion $b_0^+$ appearing in the
propagator. Since $b_0^+$ is the insertion corresponding to the modulus
that changes the length of the propagator, this term corresponds to
the new surfaces. Indeed, making use of Eqn.\pppss\ we see that it
is precisely of the same form as the second term in \wwiu.
This concludes our argument that for the case of deformations
arising from the change of stub length the ansatz for
$e_g^n$ is correct (up to combinatorial factors to be dealt
with in the next subsection).

%%%%%%%%%%%%%%%%%%%%%%%%%%%%%%%%%%%%%%%%%%%%%%%%%%%

\section{The General Case}

We now turn to a complete proof of Eqn.\lotsh. This will require
refining the definition of the vector field $\widehat U$ introduced
in \S4.2, so as to have compatibility with sewing. This is what
we do next.

Recall that for any $\V_{g,n}(u_0)$ the diffeomorphisms
$f^{\widehat U}_s$ take $\V_{g,n}(u_0)\, \rightarrow\, \V_{g,n}(u_0+s)$
and $\partial\V_{g,n}(u_0)\rightarrow\partial\V_{g,n}(u_0+s)$,
and $\widehat U$ is the vector field that generates the map.
The compatibility with sewing is the requirement that the map
$\partial\V_{g,n}(u_0)\rightarrow\partial\V_{g,n}(u_0+s)$ must
be special. Since every surface in $\partial\V_{g,n}$ is obtained
by sewing of two distinct surfaces, or of two punctures in a single
surface, we demand that the map should take the sewn surface to the
surface obtained by sewing the deformed constituents.

In order to construct this map we proceed by induction, increasing in
each step the dimensionality of the moduli spaces.
The idea will be the following: a surface on the boundary of a subspace
$\V (u_0)$ determines via Eqn.\gccon\ either a pair of surfaces
(or a single surface) in subspaces of lower dimensionality.
If we know how to map such subspaces, this determines two surfaces
(or a single one) in the new sections, which by sewing will also
determine a surface in the boundary of $\V (u_0 +s)$.
The map is then extended to the interior arbitrarily. What follows
is a detailed construction of this map. The argument is somewhat
technical as it uses the language of fiber bundles, and will not
be required for the later arguments.

\subsection{Construction of the Map}
In doing our construction we will assume that each of the terms on
the right hand side of the geometrical equation \gccon\ gives
a disjoint contribution to the boundary of $\V_{g,n}$, that is,
there is no possible cancellation of terms on the right hand side.
It is then clear that the space $\partial\V_{g,n}$ has the structure of
the sum of spaces each of which is a $U(1)$ fiber bundle with
base space $\V_{g_1,n_1}\times \V_{g_2,n_2}$ or $\V_{g-1,n+2}$,
where the $U(1)$ fiber arises from the twist operation necessary when
sewing. The map
$f_s^{\widehat U}$ discussed in \S4.2 is a diffeomorphism
taking the corresponding bundles into each other, but it need
not be fiber preserving. The compatibility with sewing
requires modifying the the diffeomorphism making it fiber preserving.

For the unique dimension zero moduli space $\V_{0,3}$ we simply define
$f_s^{\widehat U} : \V_{0,3}(u_0) \rightarrow \V_{0,3}(u_0+s)$.
This satisfies all the requirements. The
induction hypothesis is that the desired map has been defined for
all $g',n'$ such that the dimensionality of the respective subspaces
is smaller than that of $\V_{g,n}$.  We must then show how to construct
the map for $\V_{g,n}$.

To do this the most nontrivial step is to construct a diffeomorphism
mapping the various boundary subspaces $\partial\V_{g,n}(u)$.
In order to construct this map we
begin by defining a diffeomorphism
between the {\it base manifolds} of the $U(1)$ fiber bundles
representing $\partial\V_{g,n}(u)$.
Pick a surface $\Sigma \in \partial\V_{g,n} (u_0)$.
This determines via \gccon\ either two surfaces
$\widehat \Sigma_1 \in \V_{g_1,n_1} (u_0)$,
$\widehat\Sigma_2 \in \V_{g_2,n_2} (u_0)$, or a single surface
$\widehat \Sigma_l \in \V_{g-1,n+2} (u_0)$. The pair of surfaces
$(\widehat\Sigma_1 , \widehat\Sigma_2 ) \in \V_{g_1,n_1}(u_0)\times
\V_{g_2,n_2}(u_0)$ represents a basepoint, and the set of surfaces
obtained by sewing this pair of surfaces and twisting is the fiber
(similarly for $\widehat\Sigma_l$).
Now we can use the diffeomorphisms available in
the lower dimensional subspaces to define
$$\widehat\Sigma_i (u_0 +s) \equiv
f_s^{\widehat U}( \widehat\Sigma_i ) ,
\quad \hbox{for}\, \, i =1,2, \,\, \hbox{or}\,\, i = l.\eqn\dmcs$$
These surfaces, by definition, lie on the subspaces at parameter
value $(u_0 +s)$. They define a basepoint on the fiber bundles
$\partial\V_{g,n}(u_0+s)$ since sewing them gives a
specific fiber. This is a diffeomorphism between the base spaces.

We must now extend this diffeomorphism to a continuous map
between the fiber bundles. This map will be fiber preserving,
and equivariant, in the sense that it commutes with the twist
operation. For this, fix the phases of the
local coordinates around the punctures to be sewn so that the sewing
parameter is $t=1$. Let $\Sigma (\theta)\in \partial\V_{g,n}(u_0)$
be the surface obtained sewing $\widehat\Sigma_1$ and $\widehat\Sigma_2$
(or $\widehat\Sigma_l$) with $t=\exp (i\theta )$. Clearly
$\Sigma (\theta=0) = \Sigma$. Again, we deform the constituents, but
in order to be able to sew back unambiguously we have to keep track
of the phases around the punctures.
To that effect we choose a phase convention
around the punctures to be sewn for the one parameter
set of surfaces in Eqn.\dmcs. Even though
this choice cannot be done continuously for all surfaces
in the subspaces $\V$ to be sewn, this will cause no difficulty
since the continuity of the map is guaranteed by keeping track of
phases and not by how we choose them.
We can now define the map between $\partial\V_{g,n}(u_0)$
and $\partial\V_{g,n}(u_0+s)$
$$f_s^{\widehat U} : \quad \Sigma (\theta ) \quad\rightarrow\quad
\widehat\Sigma_1 (u_0+s) \cup_\theta \widehat\Sigma_2 (u_0+s)
\quad (\hbox{or} \,\, \cup_\theta\widehat\Sigma_l(u_0+s) ),\eqn\malst$$
where $\cup_\theta$ denotes sewing with $t=\exp (i\theta )$.
Due to \gccon\
it is clear that the map gives a surface in $\partial\V (u_0+s)$.
The map, by construction, is one to one (since there is no overcounting,
two surfaces built with identical constituents but different sewing
parameter cannot be the same) and onto, that is, any surface in
$\partial\V (u_0+s)$ arises from some surface on $\partial\V (u_0)$.
The map is continuous since it was
built keeping track of the phases on the punctures that were sewn, and
is equivariant with respect to $U(1)$.
Using the diffeomorphism between the base manifolds, all the fiber
bundles for various $s$ can be thought of as having the same base
space and projection given by the composition of the inverse
diffeomorphism and the original projection.
Then our continuous maps define bundle isomorphisms. But $U(1)$
bundles isomorphic in the continuous category are also isomorphic
in the smooth category \foot{We are very grateful to H. Miller
for his explanations on fiber bundles.}. Thus the map can be
made into a fiber preserving and equivariant diffeomorphism.
This is essentially what we wanted. We then obtain the vector
field $\widehat U$, defined on the boundaries $\partial\V_{g,n}(u)$
with the desired properties. Finally the map is extended smoothly
to the interior of the $\V_{g,n}$ subspaces (using the previously
defined (\S4.2) vector field $\widehat U$).
This concludes the construction
of the map at this order, and by the induction hypothesis, the desired
map and the associated generating vector field $\widehat U$ have
been shown to exist.

\subsection{Completing the Derivation}
We can now see why we expect equation \lotsh\ to hold.
The second and last terms show amplitudes constructed by first
deforming the constituent surfaces and then sewing, while the
first term shows an amplitude where we deform a sewn surface.
These two types of terms are related because
the vector field $\widehat U$ on sewn surfaces was constructed to
give the deformation induced by first
deforming the constituent surfaces and then sewing.
Let us now verify that as a result the terms cancel out precisely.

In terms of states, a surface $\Sigma \in \partial \V_{g,n}$ built
sewing, with parameter $t=\exp(i\theta)$, two surfaces
$\widehat\Sigma_1$ and $\widehat\Sigma_2$, is given by
$$\bra{\Sigma} = \bra{\widehat\Sigma_1}\bra{\widehat\Sigma_2}
R^\theta_{r_1r_2} \rangle ,
\eqn\sttws$$
where $\ket{R^\theta_{r_1r_2}}$ is the sewing ket. The key point in
our construction is that the deformation of $\Sigma$ defined
by the vector field $\widehat U$ is precisely reproduced by
deforming $\widehat\Sigma_1$ and $\widehat\Sigma_2$ with their
respectives $\widehat U$'s and then sewing
with the same sewing parameters. This means that
$$\bra{\Sigma} \bigl( 1 + \epsilon {\bf T}({\bf u} )\bigr) =
\bigl[ \bra{\widehat\Sigma_1} \bigl( 1 + \epsilon {\bf T}
({\bf u}_1) \bigr)\bigr]\,
\bigl[ \bra{\widehat\Sigma_2} \bigl( 1 + \epsilon {\bf T}
({\bf u}_2) \bigr)\bigr] \ket{R^\theta_{r_1r_2}},\eqn\rssts$$
where we used the fact that the stress tensor generates the
deformations of the states (cf. [\zwiebachlong] Eqn.(7.32)).
It follows that
$$\bra{\Sigma} {\bf T}({\bf u} ) =
\bra{\widehat\Sigma_1} \, \bra{\widehat\Sigma_2}
\bigl( {\bf T}({\bf u}_1)  + {\bf T}({\bf u}_2) \bigr)
\ket{R^\theta_{r_1r_2}} .\eqn\rsstsi$$
Since antighosts have the same connection conditions as the
energy momentum tensor (both transform as quadratic differentials)
we also have that
$$\bra{\Sigma} {\bf b}({\bf u} ) =
\bra{\widehat\Sigma_1} \, \bra{\widehat\Sigma_2}
\bigl( {\bf b}({\bf u}_1)  + {\bf b}({\bf u}_2) \bigr)
\ket{R^\theta_{r_1r_2}} .\eqn\rsstsi$$

We are going to analyze the first term in Eqn.\lotsh. In particular,
we consider the integrand evaluated on a surface obtained by sewing
two surfaces together. This case relates to the second term
of the equation, and is the only case we will discuss (the treatment
of the case when the surface is obtained by sewing two punctures of
a single surface leads to the third term in the equation; its treatment
is completely analogous). Consider then the expression
$$\eqalign{
& i_{{}_{\widehat U}} \Omega^{(0)g,n}_{\Psi^n}
\bigl( \widehat V^{\Sigma}(1,\Sigma_1),\cdots ,
\widehat V^{\Sigma}(d_1,\Sigma_1);\,
\widehat V^{\Sigma}(1,\Sigma_2),\cdots ,
\widehat V^{\Sigma}(d_2,\Sigma_2);\,
\widehat V^{\Sigma}_\theta \bigr) \cr
& \, = N_{g,n}\bra{\Sigma} {\bf b}({\bf u})
{\bf b}({\bf v}^\Sigma (1,\Sigma_1))  \cdots
{\bf b}({\bf v}^\Sigma (d_1,\Sigma_1))\,
{\bf b}({\bf v}^\Sigma (1,\Sigma_2))\cdots
{\bf b}({\bf v}^\Sigma (d_2,\Sigma_2))
{\bf b}({\bf v}^\Sigma_\theta)\ket{\Psi}^n, \cr}\eqn\bedf$$
where, following the notation of Ref.[\zwiebachlong] (\S 8),
$\widehat V^\Sigma (j,\Sigma_i)$ denotes the deformation induced on
the sewn surface $\Sigma$ by the $j$-th deformation
$\widehat V^{\Sigma_i}_j$ of the constituent surface $\Sigma_i$.
Using now the sewing representation of the surface $\Sigma$
in Eqn.\sttws , along with Eqn.\rsstsi\ and the
analogous equations for the other deformations (cf. [\zwiebachlong]
\S 8), we get that the above is given by
$$ N_{g,n}\bra{\widehat\Sigma_1}\bra{\widehat\Sigma_2}
\bigl( {\bf b}({\bf u}_1) + {\bf b}({\bf u}_2) \bigr)
{\bf b}({\bf v}^{\Sigma_1}_1)  \cdots
{\bf b}({\bf v}^{\Sigma_1}_{d_1})\,
{\bf b}({\bf v}^{\Sigma_2}_1) \cdots
{\bf b}({\bf v}^{\Sigma_2}_{d_2})\,
i(b_0 - \overline b_0)^{(r_1)}\ket{R^\theta_{r_1r_2}}
\ket{\Psi}^n .\eqn\bexcdf$$
Using now the explicit expression for the sewing ket ([\zwiebachlong],
Eqn.(2.74)) we have
$$i(b_0 - \overline b_0)^{(r_1)}\ket{R^\theta_{r_1r_2}}
= (2\pi i) \sum_s (-)^{\Phi_s}\, \ket{\widetilde \Phi_s}_{(r_1)}
{1\over 2\pi} \exp(i\theta L_0^-) \ket{\Phi_s}_{(r_2)},\eqn\bupref$$
and \bexcdf\ can then be rewritten as
$$\eqalign{
(2\pi i N_{g,n}) \, \sum_s (-)^{\Phi_s} \biggl(
&\,\bra{\widehat\Sigma_1} {\bf b}({\bf u}_1)
{\bf b}({\bf v}^{\Sigma_1}_1)  \cdots
{\bf b}({\bf v}^{\Sigma_1}_{d_1})
\ket{\Psi}^{n_1-1}\ket{\widetilde\Phi_s}_{(r_1)}\cr
&\,\cdot \bra{\widehat\Sigma_2} {\bf b}({\bf v}^{\Sigma_2}_1)  \cdots
{\bf b}({\bf v}^{\Sigma_2}_{d_2}){1\over 2\pi} \exp (i\theta L_0^-)
\ket{\Phi_s}_{(r_2)} \ket{\Psi}^{n_2-1}\cr
&+(-)^{\Phi_s+1}
\bra{\widehat\Sigma_1} {\bf b}({\bf v}^{\Sigma_1}_1)  \cdots
{\bf b}({\bf v}^{\Sigma_1}_{d_1})
\ket{\Psi}^{n_1-1}\ket{\widetilde\Phi_s}_{(r_1)}\cr
&\,\cdot\bra{\widehat\Sigma_2}{\bf b}({\bf u}_2)
{\bf b}({\bf v}^{\Sigma_2}_1)  \cdots
{\bf b}({\bf v}^{\Sigma_2}_{d_2}){1\over 2\pi} \exp (i\theta L_0^-)
\ket{\Phi_s}_{(r_2)} \ket{\Psi}^{n_2-1}\, \biggr). \cr}\eqn\ffack$$
At the level of forms this just means that
$$\eqalign{
& i_{{}_{\widehat U}} \Omega^{(0)g,n}_{\Psi^n}
\bigl( \widehat V^{\Sigma}(1,\Sigma_1),\cdots ,
\widehat V^{\Sigma}(d_1,\Sigma_1);\,
\widehat V^{\Sigma}(1,\Sigma_2),\cdots ,
\widehat V^{\Sigma}(d_2,\Sigma_2);\,
\widehat V^{\Sigma}_\theta \bigr)\cr
&= \sum_s (-)^{\Phi_s}\biggl(
i_{{}_{\widehat U}} \Omega^{(+1)g_1,n_1}_{\Psi^{n_1-1}\widetilde\Phi_s}
(\widehat V^{\Sigma_1}_1,\cdots ,\widehat V^{\Sigma_1}_{d_1} )\cdot
\Omega^{(0)g_2,n_2}_{
{1\over 2\pi}(\exp (i\theta L_0^-)\Phi_s)\Psi^{n_2-1} }
(\widehat V^{\Sigma_2}_1,\cdots ,\widehat V^{\Sigma_2}_{d_2} )\cr
&\quad\quad +(-)^{\Phi_s+1}
\Omega^{(0)g_1,n_1}_{\Psi^{n_1-1}\widetilde\Phi_s}
(\widehat V^{\Sigma_1}_1,\cdots ,\widehat V^{\Sigma_1}_{d_1} )\cdot
i_{{}_{\widehat U}}\Omega^{(+1)g_2,n_2}_{ {1\over 2\pi}
(\exp (i\theta L_0^-)\Phi_s)\Psi^{n_2-1} }
(\widehat V^{\Sigma_2}_1,\cdots ,\widehat V^{\Sigma_2}_{d_2} )\,
\biggr) ,\cr}\eqn\atherx$$
where we have used that $2\pi iN_{g,n}=N_{g_1,n_1}N_{g_2,n_2}$.
Integrating over the sewing angle and using Eqn.(2.83) of
Ref.[\zwiebachlong] on the second term, we obtain
$$\eqalign{
& \int\!\!d\theta\,i_{{}_{\widehat U}} \Omega^{(0)g,n}_{\Psi^n}
\bigl( \widehat V^{\Sigma}(1,\Sigma_1),\cdots ,
\widehat V^{\Sigma}(d_1,\Sigma_1);\,
\widehat V^{\Sigma}(1,\Sigma_2),\cdots ,
\widehat V^{\Sigma}(d_2,\Sigma_2);\,\widehat V^{\Sigma}_\theta)\cr
&= {\sum_s}' (-)^{\Phi_s}\biggl(
i_{{}_{\widehat U}} \Omega^{(+1)g_1,n_1}_{\Psi^{n_1-1}\widetilde\Phi_s}
(\widehat V^{\Sigma_1}_1,\cdots ,\widehat V^{\Sigma_1}_{d_1} )\cdot
\Omega^{(0)g_2,n_2}_{\Phi_s\Psi^{n_2-1} }
(\widehat V^{\Sigma_2}_1,\cdots ,\widehat V^{\Sigma_2}_{d_2} )\cr
&\quad\quad +i_{{}_{\widehat U}}
\Omega^{(+1)g_2,n_2}_{\Psi^{n_2-1}\widetilde\Phi_s }
(\widehat V^{\Sigma_2}_1,\cdots ,\widehat V^{\Sigma_2}_{d_2} )
\cdot\Omega^{(0)g_1,n_1}_{\Phi_s\Psi^{n_1-1}}
(\widehat V^{\Sigma_1}_1,\cdots ,\widehat V^{\Sigma_1}_{d_1} )
\,\biggr) .\cr}\eqn\atherx$$
We now simply recall that
$$\int_{(\partial \V_{g,n})_{\scriptstyle \rm tree}} = -{1\over 2}\,
\sum\limits_{{g_1+g_2=g}\atop {n_1+n_2 = n+2}}
\int_{\V_{g_2,n_2}}\cdot\int_{\V_{g_1,n_1}}\cdot\int d\theta
\cdot {n! \over (n_1-1)!(n_2-1)!} , \eqn\tleqben$$
where the combinatorial factor arises because there are
that number of ways of splitting $n$ string fields into two
subsets with $(n_1-1)$ and $(n_2-1)$ string fields each.
The last two equations imply the desired cancellation
in Eqn.\lotsh\ between
the part of the first term having to do with surfaces sewn in the
tree configuration and the second term. The third term cancels against
the part of the first term having to do with surfaces sewn in the
loop configuration, by a completely analogous argument. This proves the
validity of Eqn.\lotsh , and as a consequence concludes our
proof that the quoted infinitesimal parameter indeed reproduces
the change in the action induced by a change in the decomposition of
moduli space.

%%%%%%%%%%%%%%%%%%%%%%%%%%%%%%%%%%%%%%%%%%%%%%%%%%%%%%%%%%%%%%%%%%%%%

\chapter{The Hamiltonian Vector Field for the Quantum Master Action}

Given a string action $S$, that is, a function on the supersymplectic
manifold $\M$, we denote the corresponding hamiltonian vector field by
$V_S$. By definition it is given as $i_{{}_{V_S}} \omega = -\hbox{d} S$,
or, more explicitly, by Eqn.\HV. Witten showed [\witten] that
the vector $V_S$ corresponding to an action that satisfies the
classical master equation must satisfy $V_S^2 =0$. We now study the
generalization for the case of the full master equation.

\REF\stasheff{T. Lada and J. Stasheff, `Introduction to sh Lie algebras
for physicists', UNC-MATH-92/2, hep-th 9209099, September 1992.}

The master equation for the action demands that
$$ \hbar \Delta S + {1\over 2}\, \{ S , S \} = 0, \quad
\Delta S \equiv {1\over 2} \hbox{div} \, V_S . \eqn\meaction$$
The left hand side of the master equation is a function that must
vanish. Therefore it follows that the hamiltonian vector corresponding
to this function must vanish too:
$$ V_{\hbar\Delta S + {1\over 2}\, \{ S , S \} } = 0 .\eqn\mustvan$$
It follows from linearity that
$$ \hbar V_{\Delta S} + {1\over 2}\,V_{ \{ S , S \} } = 0 .
\eqn\mustvani$$
Using the basic relation between (super)Lie brackets and
(anti)brackets, Eqn.\ALGEBRA,
and the definition of $\Delta$ in Eqn.\meaction , we find
$$ V_S^2 = -{\hbar\over 2}  V_{\hbox{div} V_S} ,\eqn\resultf$$
which is the generalization of the equation $V_S^2 = 0$ given in
[\witten] necessary for the action $S$ to satisfy the full master
equation. While this equation is necessary for the master equation to
hold, it is not quite sufficient to guarantee it.
If Eqn.\resultf\  holds, then the hamiltonian
vector field in \mustvan\  does indeed vanish. Given a vanishing
hamiltonian vector field, the corresponding hamiltonian function
$$ \hbar \Delta S + {1\over 2}\, \{ S , S \} =
\Half \bigl( \hbar \hbox{div} V_S + \omega (V_S ,V_S)\bigr) ,
\eqn\divan$$
could be a constant different from zero. The way to argue that this
constant is zero is to show that the vector field $V_S$ solving
\resultf\ has at least one zero in the supermanifold. This would make
the right hand side of the above equation vanish at that point,
and therefore everywhere. In the language
of homotopy Lie algebras \foot{See Ref.[\wittenzwiebach], and
Ref.[\stasheff ] for an introduction to the basic concepts.}
for such a vector would look like
$$V_S = \bigl( f_{a_1}^b \eta^{a_1} + f_{a_1a_2}^b \eta^{a_1}\eta^{a_2}
+ \cdots \bigr) {\partial\over \partial \eta^b},\eqn\vfwz$$
where we have a zero at $\eta^a_1 = \cdots = \eta^{a_n} = 0$.

It follows from [\zwiebachlong] that
there is a nontrivial vector field $V_S$ solving Eqn.\resultf.
Nevertheless the vector does not have a zero! In those
solutions $V_S$ reads
$$\eqalign{V_S &= \bigl(
f_{a_1}^{(0)b} \eta^{a_1} + f_{a_1a_2}^{(0)b} \eta^{a_1}\eta^{a_2}
+ \cdots \bigr) {\partial\over \partial \eta^b} \cr
&\quad + \hbar \bigl(  f^{(1)b}  + f_{a_1}^{(1)b} \eta^{a_1}
+ \cdots \bigr) {\partial\over \partial \eta^b}+ \O (\hbar^2 ),\cr}
\eqn\vxfwz$$
and therefore it is of $\O (\hbar )$ for $\eta^a=0$. The terms
$f^{(1)b}$, for example, have to do with one point functions of
arbitrary states on genus one surfaces. Since we do not have a
zero, verifying that there are solutions where \divan\ vanishes
is not trivial. Fortunately this was shown
already in Ref.[\zwiebachlong] (see around Eqn.(3.42)) where it was
shown that no constant terms can arise on the BV master equation
($=$lhs of \divan ) because of ghost number conservation.

It is clear from the above considerations that in order to
formulate the quantum closed string field theory in the space of
two-dimensional theories we need to find not only a closed
non-degenerate two-form $\omega$, but also a density $\rho$.
This density cannot be fixed arbitrarily; it must lead to
$\Delta_\rho^2=0$ (thus given $\omega$ we cannot assume that
$\rho$ is a constant). Once we find a suitable density,
we must solve Eqn.\resultf. A particularly simple solution would
involve a vector $V_S$ satisfying both $V_S^2 = 0$ {\it and}
$\hbox{div}V_S = 0$. Our experience with closed string field
theory, where the path integral measure is not invariant under
the gauge symmetry, suggests that this is requiring too much.
We expect that Eqn.\resultf\ is satisfied in the weakest form.

\ack
We are grateful to K. I. Izawa, H. Miller, K. Ranganathan,
H. Sonoda and A. Schwarz for useful discussions.
H.Hata would like to acknowledge the hospitality of Center for
Theoretical Physics at MIT, where this work was started.

%%%%%%%%%%%%%%%%%%%%%%%%%%%%%%%%%%%%%%%%%%%%%%%%%%%%%%%%%%%%%%%%%%%%%

\APPENDIX{A}{A:\ Exterior calculus on a supermanifold}

In this Appendix, we summarize the basic notions of exterior calculus
on a supermanifold. In contrast to the case of ordinary manifolds we
have two kinds of odd objects; the exterior derivative operator $\d$,
and the coordinates $z^I$ that are odd. We adopt the convention that
these two {\it commute} with each other; the Grassmanality of
$\d z^I$ is the same as that of $z^I$: $\E{\d z^I}=\E{z^I}=I$.
The wedge product is defined by
$$
\d z^{I_1}\wedge \d z^{I_2}\wedge \cdots\wedge \d z^{I_N}
= \sum_{\sigma\in S_N}\e(\sigma)\eta(\{I\},\sigma)
\d z^{I_{\sigma(1)}}\otimes \d z^{I_{\sigma(2)}}
\otimes\cdots\otimes \d z^{I_{\sigma(N)}}\ ,
\eqn\WEDGEPROD
$$
where $\e(\sigma)$ is the signature of the permutation $\sigma$,
and the extra sign factor $\eta(\{I\},\sigma)$ is that necessary
for reordering $z^{I_1}\cdots z^{I_N}$ into
$z^{I_{\sigma(1)}}\cdots z^{I_{\sigma(N)}}$.
Therefore, we have
$$\d z^I\wedge \d z^J = -(-)^{IJ}\d z^J\wedge \d z^I\ ,\eqn\DZDZ$$
and
$$\alpha\wedge\beta = (-)^{mn+\E{\alpha}\E{\beta}}\beta\wedge\alpha \ ,
\eqn\ALPHABETA$$
for a general $m$-form
$\alpha=\alpha(z)_{I_1\cdots I_m}
\d z^{I_1}\wedge\cdots\wedge\d z^{I_m}$
and a general $n$-form $\beta$. Here $\e (\alpha ) = \e (\alpha (z))
+ \sum_m \e (z^{I_m})$. The exterior derivative $\d$ acting on a
function $f(z)$ is given by
$$\d f = \d z^I\dl{I}f = f\dr{I}\d z^I\ .\eqn\DF$$

The basis of vectors arising from a local coordinate system $(z^I)$
is denoted by
$(\tv{I})_{I=1,\ldots ,2n}$, and a general vector field $V$ is
expanded as
$$
V(z) = \TV{I}V^I(z)\ .
\eqn\VECTORFIELD
$$
The Grassmanality of $\tv{I}$ is $\E{\tv{I}}=I$.
The reason why $\tv{I}$ carries a left-arrow is that a
vector field acts on a function $f(z)$ as the right-derivative:
$$V(f) \equiv f\dr{I} V^I\ .\eqn\VF$$
On the other hand, a one-form $\d z^I$ acts on a vector field as
$$\d z^I\biggl(\TV{J}\biggr) \equiv \delta^I_J\ .\eqn\DZV$$
More generally, for vector fields $V_i=(\tv{I})V_i^I$ we have
$$
\bigl(\d z^{I_1}\otimes\cdots\otimes \d z^{I_N}\bigr)(V_1,\cdots,V_N)
=(-)^E \d z^{I_1}(V_1)\cdots \d z^{I_N}(V_N)
= (-)^E V_1^{I_1}\cdots V_N^{I_N}\ ,
\eqn\DZDZVV
$$
where $E=\sum_{i=1}^{N-1}\E{V_i}\sum_{j=i+1}^N\E{I_j}$.

We now summarize a number of properties and operations on a general
$N$-form $\Omega$. First, we have
$$
\Omega(V_1,\cdots,V_i,V_{i+1},\cdots,V_N) =
- (-)^{V_i V_{i+1}}\Omega(V_1,\cdots,V_{i+1},V_i,\cdots,V_N)\ .
\eqn\EXCHG
$$
Next we define the sign exponents $R_i^{(N)}$ and $L_{ij}^{(N)}$
which arise in reordering the vector fields $V_i$ due to their
Grassmanality by
$$
\eqalign{
V_1V_2\cdots V_N &= (-)^{R_i^{(N)}}V_1\cdots\check V_i\cdots V_N V_i \cr
&= (-)^{L_{ij}^{(N)}}V_i V_j V_1\cdots
\check V_i\cdots\check V_j\cdots V_N\ , \cr
}
\eqn\DEFRL
$$
where the check on $\check V$ implies the omission of that $V$.
Then the exterior derivative $\d$,
the contraction operator $i_W$, and the Lie-derivative
$\L_V\equiv \d i_V + i_V\d$ are given
as follows:
$$
\eqalignno{
&\d\Omega\left(V_1,V_2,\cdots,V_{N+1}\right) =
\sum_{i=1}^{N+1} (-)^{i+1+R_i^{(N+1)}}
\Omega(V_1,\cdots,{\check V}_i,\cdots,V_{N+1})V_i \cr
&\quad -\sum_{1\le i<j\le N+1}
(-)^{i+j+L_{ij}^{(N+1)}}
\Omega([V_i,V_j],V_1,\cdots,{\check V}_i,\cdots,
{\check V}_j,\cdots,V_{N+1}) \ ,
&\eqname\DOMEGA \cr
\noalign{\vskip.5cm}
&\left(i_W \Omega\right)(V_1,\cdots,V_{N-1})
= \Omega(W, V_1,\cdots,V_{N-1}) \ ,
&\eqname\CONTRACTION \cr
\noalign{\vskip.5cm}
&\left(\L_W\Omega\right)(V_1,\cdots,V_N)
= (-)^{W\sum_{i=1}^N V_i}\Omega(V_1,\cdots,V_N)W \cr
&\qquad\qquad\qquad\qquad\qquad
+ \sum_{i=1}^N (-)^{W\sum_{j=1}^{i-1}V_j}
\Omega(V_1,\cdots,[W,V_i],\cdots,V_N) \ .
&\eqname\LIEDERIVATIVE \cr}$$
In Eqn.\DOMEGA, $V_i$ in the second term acts on the function
$\Omega(V)$ on the left.
Another useful formula is
$$[\L_V,i_W] \equiv \L_V i_W - (-)^{VW}i_W\L_V = i_{[V,W]}\ .\eqn\LIEQI$$

%%%%%%%%%%%%%%%%%%%%%%%%%%%%%%%%%%%%%%%%%%%%%%%%%%%%%%%%%%%%%%%%%%%%%
\APPENDIX{B}{B:\ Properties of various quantities}

We denote by $\E{\O}$ the Grassmanality of the quantity \O:
$\E{\O}=0$ ($1$) (mod 2) when \O\ is Grassmann even (odd).
We also use the abbreviations
$\E{z^I}=I$ and $(-)^{\E{\O}}=(-)^{\O}$.

\noindent
1) The right and left-derivatives with respect to $z^I$ acting on
a general quantity $A(z)$ are related by
$$
A\dr{I} = (-)^{I(A+1)}\dl{I}A \ .
\eqn\RLDERIV
$$

\noindent
2) $\omega_{IJ}$ and $\omega^{IJ}$:
$$
\eqalign{
&\E{\omega_{IJ}} = \E{\omega^{IJ}} = I+J+1 \ , \cr
&\omega_{IJ} = -(-)^{IJ}\omega_{JI} \ , \cr
&\omega^{IJ} = -(-)^{(I+1)(J+1)}\omega^{JI} \ . \cr
}\eqn\PROPOMEGA
$$

\noindent
3) Antibracket:
$$
\eqalign{
&\E{\{A,B\}} = A + B + 1 \ , \cr
&\{A,B\}= - (-)^{(A+1)(B+1)}\{B,A\} \ , \cr
&\{A, BC\} = \{A,B\}C + (-)^{(A+1)B}B\{A,C\} \ , \cr
&\{AB, C\} = A\{B,C\} + (-)^{B(C+1)}\{A,C\}B \ .\cr
}
\eqn\PROPAB
$$

\noindent
4) The Jacobi identity \JACOBI\ in terms of
$\omega_{IJ}$ and $\omega^{IJ}$:
$$
\eqalign{
&(-)^{IK}\p_I \omega_{JK} + \hbox{cyclic}(I,J,K) = 0 \ , \cr
&(-)^{(I+1)(K+1)}\omega^{IL}\p_L \omega^{JK}
+ \hbox{cyclic}(I,J,K) = 0 \ . \cr
}
\eqn\JACOBICOMP
$$

%%%%%%%%%%%%%%%%%%%%%%%%%%%%%%%%%%%%%%%%%%%%%%%%%%%%%%%%%%%%%%%%%%%%%
\APPENDIX{C}{C:\ Finite transformations}

Given a canonical diffeomorphism $g: \M \rightarrow \M$, one
has $g^*\omega = \omega$ and
$$\eqalignno{
&\{g^*A,g^*B\}=g^*\{A,B\} ,\qquad g_* V_{g^*A} = V_A, &\eqname\CTII \cr
\noalign{\hbox{and}}
&\wt{\omega}^{IJ}(z) = \omega^{IJ}(z) , &\eqname\CTIII \cr}$$
where $\bigl(g^* A\bigr)(z)=A\bigl(g(z)\bigr)$ for a 0-form $A(z)$,
and $\wt{\omega}^{IJ}$ is defined by Eqn.\OMEGATILDE\ with
$\wt{z}^I = g^I(z)$.
The invariance, $\A(S^g)=\A(S)$, is proven as follows.
Defining $H(z) \equiv \exp S(z)$, for which the transformation
\FINITEDE\ reads $H\rightarrow (g^*H)\bigcdot F_g$,
we have
$$\eqalign{
&\A(S^g) = -\Half\int_\M\!d\mu \{(g^*H)F_g,(g^*H)F_g\} \cr
&=-\Half\int_\M\! d\mu\left(
F_g^2\{g^*H,g^*H\} + \Half\{(g^*H)^2,F_g^2\}
+ (g^*H)^2\{F_g,F_g\}\right) \cr
&= -\Half\int_\M\! d\mu\,F_g^2\,g^*\!\left(\{H,H\}\right)
+ \Half\int_\M\! d\mu\,(g^*H)^2
\left(\Delta F_g^2 - \{F_g,F_g\}\right) , \cr}\eqn\CTAS$$
where in the last step we have used Eqns.\CTII\ and \saddel.
The first term in the last expression of Eqn.\CTAS\ is in fact
equal to $\A(S)$
$$
-\Half\int_\M\! d\mu\,F_g^2\,g^*\!\left(\{H,H\}\right)
= -\Half\int_\M\! g^*\!\left(d\mu\,\{H,H\}\right)
=\A(S) ,
$$
where use was made of Eqn.\rmeas.
Therefore, $\A(S^g)=\A(S)$ holds if
$\Delta F_g^2 - \{F_g,F_g\} =0$. Using Eqn.\ABDELTA , we must
show that
$$\Delta_\rho F_g = 0 ,\eqn\CTCOND$$
where we have made explicit the $\rho$ dependence of $\Delta$.
Since the delta operator is a scalar under coordinate transformations,
and our transformations leave $\omega$ invariant, we have
$$
\Delta_\rho = \Delta_{\wt{\rho}}\vert_{z\rightarrow\wt{z}} ,
\eqn\TWODELTAS
$$
for $\wt{\rho}$ defined by Eqn.\RHOTILDE,
and hence the nilpotency of $\Delta_\rho$ implies the nilpotency of
$\Delta_{\wt{\rho}}$.
{}From this fact and the formula [\schwarz]
$$
\Delta_{\wt{\rho}}^2 A = \Delta_\rho^2 A +
2\{e^{-\sigma/2}\Delta_\rho e^{\sigma/2}, A\}\ ,
\eqn\DELTAWTRHOSQUARE
$$
where $\sigma =\ln\left(\wt{\rho}/\rho \right)$,
it follows that
$$
\Delta_\rho e^{\sigma/2} = 0 .
\eqn\SCHWARZFML
$$
The last equation also implies that
$$
\Delta_{\wt{\rho}} e^{-\sigma/2} = 0 .
\eqn\SCHWARZFMLII
$$
Eqn.\CTCOND\ is obtained from Eqn.\SCHWARZFMLII\ by making the
replacement $z\rightarrow\wt{z}=g(z)$, using Eqn.\TWODELTAS\ and the
fact that $F_g(z) = \exp\left(-\sigma(g(z))/2\right)$.
This finishes the proof of the invariance of $\A(S)$ under the
transformation \FINITEDE.
%%%%%%%%%%%%%%%%%%%%%%%%%%%%%%%%%%%%%%%%%%%%%%%%%%%%%%%%%%%%%%%%

\refout
\figout
\end